\documentclass[11pt]{article}
\usepackage{caption}
\usepackage{scalerel}
\usepackage{amsthm,latexsym,amsxtra,graphicx,epstopdf,hyperref,setspace,fix-cm}
\usepackage[normalem]{ulem}
\usepackage{makeidx}
\usepackage{fullpage}
\usepackage{amsmath}
\usepackage{amssymb}
\usepackage{setspace}
\usepackage{bbm}
\usepackage{dsfont}
\usepackage{graphics}
\usepackage{epsfig}
\usepackage[font=footnotesize,labelfont=bf,justification=centerlast,width=.94\textwidth]{caption}
\usepackage{cite}
 \usepackage{multirow}
\usepackage{array,booktabs}
\usepackage{subfigure}
\usepackage{color}
\usepackage[makeroom]{cancel}
\usepackage{pgfplots}
\usetikzlibrary{intersections, pgfplots.fillbetween}
\usepackage{tensor}
\usepackage{simplewick}

\usepackage{caption}
\usepackage{hyperref}

\hypersetup{
    colorlinks,%
    citecolor=blue,
    filecolor=blue,%
    linkcolor=blue,%
    urlcolor=blue
}
\usepackage{float}
\usepackage{slashed}
\usepackage{tikz}
\usetikzlibrary{decorations.pathmorphing}

\renewcommand{\appendixname}{}

\newcommand{\bi}{\begin{itemize}}
\newcommand{\ei}{\end{itemize}}
\newcommand{\bea}{\begin{eqnarray}}
\newcommand{\eea}{\end{eqnarray}}
\newcommand{\be}{\begin{equation}}
\newcommand{\ee}{\end{equation}}

\def\XXint#1#2#3{{\setbox0=\hbox{$#1{#2#3}{\int}$}
     \vcenter{\hbox{$#2#3$}}\kern-.5\wd0}}

\numberwithin{equation}{section}
\allowdisplaybreaks  

\usepackage{xcolor}

\newcommand{\eg}{{\it e.g.,}\ }

\newcommand{\s}{\mathfrak{s}}

\pgfplotsset{compat=1.18}

\begin{document}
\vspace*{2.5cm}
\begin{center}
\Large \textsc{Krylov complexity and chaos in deformed SYK models} \\ \vspace*{1.5cm}
\end{center}

\begin{center}
Shira Chapman${}^a$, Saskia Demulder${}^a$, Dami\'an A. Galante${}^b$, Sameer U. Sheorey${}^b$, Osher Shoval${}^a$\\ 
\end{center}

\begin{center}
{\textsf{\scriptsize{
schapman@bgu.ac.il, demulder@post.bgu.ac.il, damian.galante@kcl.ac.uk, sameer.sheorey@kcl.ac.uk, oshersho@post.bgu.ac.il}} } 
\end{center}

\begin{center}
{
\footnotesize
${}^a$Department of Physics, Ben-Gurion University of the Negev,\\ David Ben Gurion Boulevard 1, Beer Sheva 84105, Israel\\ \vspace{10pt}
${}^b$Department of Mathematics, King's College London,\\ \vspace{-2pt} Strand, London WC2R 2LS, UK
}
\end{center}

\vspace*{0.5cm}

\vspace*{1.5cm}
\begin{abstract}

\noindent Krylov complexity has recently been proposed as a quantum probe of chaos. The Krylov exponent characterising the exponential growth of Krylov complexity is conjectured to upper-bound the Lyapunov exponent. We compute the Krylov and the Lyapunov exponents in the Sachdev-Ye-Kitaev model and in some of its deformations. We do this analysis both at infinite and finite temperatures, in models where the number of fermionic interactions is both finite and infinite. We consider deformations that interpolate between two regions of near-maximal chaos and deformations that become nearly-integrable at low temperatures. In all cases, we find that the Krylov exponent upper-bounds the Lyapunov one. However, 
we find that while the Lyapunov exponent can have non-monotonic behaviour as a function of temperature, in all studied examples the Krylov exponent behaves monotonically. For instance, we find models where the Lyapunov exponent goes to zero at low temperatures, while the Krylov exponent saturates to its maximal bound. We speculate on the possibility that this monotonicity might be a generic feature of the Krylov exponent  in quantum systems evolving under unitary evolution.

\end{abstract}

\newpage

\setcounter{tocdepth}{2}
\tableofcontents

\section{Introduction} \label{sec:intro}

Dynamical probes of quantum many-body systems are 
typically hard to compute, in part due to the fact that the Hilbert space grows exponentially with the number of constituents. This difficulty is even more pronounced in quantum systems that are strongly coupled and/or quantum chaotic. Despite  significant progress in the area \cite{Haake:2010fgh},   defining a good quantum signature of chaos valid for all  time scales can be difficult. 

The situation improves when the quantum system has some intrinsically small parameter, such as in large-$N$ theories, where it is possible to take a semi-classical limit. A landmark of classical chaos is the exponential growth of the Poisson bracket as a function of time \cite{CookeRoger2005AMLt}. One can think then, that in a quantum theory with a semi-classical limit, the commutator of simple (local) operators might be related to quantum chaos. It was actually the square of the commutator \cite{Larkin} that inspired the consideration of the four-point out-of-time-ordered correlator (OTOC) at finite inverse temperature $\beta$ as a diagnosis of chaos in quantum systems. 
In a quantum chaotic system with a small parameter, that we will take to be $1/N$, this correlator generically behaves as\footnote{The formal definition of the regularized OTOC is given in  \eqref{regularised OTOC}.} 
\begin{eqnarray}
    \text{OTOC} = f_0 - \frac{f_1}{N} \exp \lambda_L t + \cdots \,,
\end{eqnarray}
where $f_{0,1}$ are order one numbers that are theory-dependent and $\lambda_L$ is known as the Lyapunov exponent. Note that this expansion defines a timescale, that we usually call the scrambling time $t_{sc} \sim \log N/\lambda_L$, at which the $1/N$ correction becomes order one. For the Lyapunov exponent to be meaningful, this time scale has to be much larger than the dissipation time $t_d\sim \beta$, which is guaranteed as long as $N\gg 1$ with $\beta\sim O(1)$. Furthermore, it was shown in \cite{Maldacena:2015waa} that in local, unitary quantum systems, $f_{0,1} >0$ and
\begin{equation}\label{BCHAOS}
\lambda_L \leq \frac{2\pi}{\beta},
\end{equation}
which we will refer to as the chaos bound. There are, of course, other interesting quantities to diagnose quantum chaos, such as the onset of random matrix statistics \cite{DAlessio:2015qtq}, the spectral form factor \cite{Stockmann_1999} or quantum circuit complexity \cite{Susskind:2018pmk,Chapman:2021jbh}.

Another related quantum signature of chaos comes from the idea of operator growth. The intuition is that, if the Hamiltonian is chaotic,  an initially simple operator will grow exponentially in complexity under Heisenberg time evolution. As opposed to the OTOC, this concept does not require a large $N$ parameter. Recently, the use of Krylov subspace methods has been proposed to quantify this idea of operator growth \cite{Parker:2018yvk}. 

Since their conception in 1931, Krylov methods have played an important role in mathematics and theoretical physics \cite{Kry31}. Their main feature is to project a higher-dimensional (computationally hard) problem to a lower-dimensional one (approximate, but computationally more accessible). Typical examples involve matrix diagonalisation or eigenvalue problems. In the case of operator growth, the idea is to find the minimal subspace needed to follow the time-evolution of an operator, without the need of diagonalising the full Hamiltonian. This is done by considering the subspace spanned by the set of nested commutators of the operator with the Hamiltonian. The nested commutators are said to form a basis of the Krylov subspace. It was conjectured that the properties of this subspace can be used to diagnose whether the quantum system is chaotic or not  by using the \emph{operator growth hypothesis} \cite{Parker:2018yvk}. Since then a critical amount of work has been devoted to understand and test this hypothesis in different quantum systems. See \cite{Nandy:2024htc}, and references therein, for a comprehensive review of recent results on the subject.

As a quantitative measure of operator growth, the concept of Krylov complexity was coined in \cite{Parker:2018yvk}. Under certain assumptions that we will discuss later, this complexity grows exponentially with time in chaotic systems, and it was conjectured that the exponent, called the Krylov exponent $\lambda_K$, would serve as a tighter bound for the Lyapunov exponent (when this one is well-defined), namely, 
\begin{equation}
\label{bound}
\mathcal{\lambda}_{L}\leq \lambda_K \leq \frac{2\pi}{\beta} ~.
\end{equation}
As we will see, while the Lyapunov exponent appears in the four-point OTOC, the computation of the Krylov exponent only requires knowledge of the two-point function of the operator. Then, if the left bound turns out to be tight, this gives a computational advantage to $\lambda_K$ as a probe of quantum chaos. It is worth noting that Krylov complexity was generalised to study the spread of states 
rather than operators both in unitary 
\cite{Caputa:2021sib,Balasubramanian:2022tpr,Caputa:2023vyr,Caputa:2024vrn} and non-unitary systems \cite{Bhattacharya:2023zqt,Bhattacharya:2023yec,Bhattacharya:2024hto}.

In this work we explore the relation between the Lyapunov exponent $\lambda_L$ and the Krylov exponent $\lambda_K$ in the Sachdev-Ye-Kitaev (SYK) model \cite{Sachdev:1992fk, kitaev_vid} and some if its deformations. The SYK model is a quantum system of $N$ Majorana fermions, with all-to-all interactions and disordered couplings, whose generalizations were proposed as a possible models for strange metallic behavior \cite{Chowdhury:2021qpy}. 
In most cases, the model develops a chaotic behaviour in its strongly-coupled phase, but it can nevertheless be solved in different limits using different numerical and analytical techniques at any temperature. In fact, the SYK model is maximally chaotic at low temperatures in the sense of \eqref{bound}. The Lyapunov exponent \cite{kitaev_vid, Maldacena:2016hyu}, the spectral form factor \cite{Cotler:2016fpe} and  operator size distribution  \cite{Roberts:2018mnp} were among the first ideas explored regarding the chaotic nature of the SYK model. Moreover, holographic black holes also saturate the bound on chaos \eqref{BCHAOS}. This surged interest in using  the SYK model as a toy model for two-dimensional holography, see \eg \cite{Sarosi:2017ykf,Rosenhaus:2018dtp}. Of particular interest to our work is the possibility of studying holographic proposals for quantum complexity from a microscopic perspective, see \eg \cite{Rabinovici:2023yex}.

We will focus on the large-$N$ limit of the SYK model and its deformations. In this limit, the model is dominated by a saddle-point approximation of its partition function.
If one then takes a further limit in which the number of fermions in each interaction term $q$ is large (but still small, compared to $N$), the system simplifies considerably. 
In this limit, the Krylov exponent satisfies $\lambda_L = \lambda_K$, both at infinite and finite temperature \cite{Parker:2018yvk}. Thus the lower bound on $\lambda_K$ in eq.~\eqref{bound} is saturated, fuelling the hope that Krylov complexity provides a computational advantage to diagnose chaos in quantum systems. We will see in this paper however that this is only the case when the SYK model is not deformed and in the strict large $q$ limit. As a first step, we extend the previous result to the next order in the large-$q$ expansion, as well as to finite $q$ and conclude that the Krylov exponent still provides a  
rather tight upper bound on the Lyapunov exponent at all temperatures. In these cases, the differences between the Krylov and the Lyapunov exponents, while non-vanishing, remain small. When $q=2$, the interaction term in the Hamiltonian becomes a random mass term, which is known to be integrable. In this case, we show that the Lanczos coefficients saturate both at finite and infinite temperature, which gives $\lambda_K = 0$.

But interestingly, the SYK model also admits a rich landscape of relevant deformations that take the form of another SYK Hamiltonian with fewer fermions in each interaction term \cite{Garcia-Garcia:2017bkg, Kim:2020mho, Garcia-Garcia:2020dzm, Lunkin:2020tbq, Nandy:2022hcm,Menzler:2024atb,Jiang:2019pam,Anninos:2020cwo, Anninos:2022qgy,Louw:2023lpq}. These deformed models can have non-trivial behaviour in the infrared, including the possibility of transitions between different regions of near-maximal chaos or transitions between near-maximally chaotic and integrable behaviour. In both cases, both the Lyapunov and the Krylov exponent can be computed and compared. We find strong evidence analysing different kinds of deformed SYK models, both analytically and numerically, that while the Lyapunov exponent can have non-monotonic behaviour as a function of the inverse temperature, the Krylov exponent can only behave monotonically. 
So, for instance, if an initially chaotic system becomes integrable (or the Lyapunov exponent becomes very small) at large inverse temperatures, $\lambda_K$ will not provide a good diagnostic of this transition. In fact, in some of the models we studied, the Lyapunov exponent decays to zero at large inverse temperatures, while the Krylov exponent saturates to its maximal value. We conjecture that, when it is well-defined, the monotonicity of the Krylov exponent may be a generic feature of unitary quantum systems.

The rest of the paper is organised as follows. In section \ref{sec:preliminaries}, we give an overview of the SYK model and its deformations, as well as of the basics of Krylov complexity. In section \ref{sec:singleSYK}, we analyse the undeformed SYK model with different values of $q$ using Krylov complexity and compare the results to those obtained from the Lyapunov exponent. Section \ref{sec:def_SYK} is devoted to study the deformed SYK models, both at infinite and finite $q$, with both integrable and chaotic deformations. Finally, in section \ref{sec:outlook}, we summarize our results, elaborate on the conjecture regarding the monotonicity of the Krylov exponent and give some future directions. A number of technical details are relegated to appendices, where, in particular, we give a detailed account on the numerical techniques used to compute the Lyapunov exponent and the spectral function in the models analysed.


\section{Preliminaries} \label{sec:preliminaries}

\subsection{Review of (deformed) SYK models}
The SYK model is a quantum mechanical model  with random, $q$-local interactions. See \cite{ Sarosi:2017ykf,Maldacena:2016hyu,Rosenhaus:2018dtp,Trunin:2020vwy,Chowdhury:2021qpy}, for instance, for some of the reviews on the model. 
The observables of the theory are built from $N$ Majorana fermions, $\psi_{i}$, obeying equal time anti-commutation relations
\begin{equation}
\label{Majorana anticom relns}
    \{\psi_i,\psi_j\} = \delta_{ij}\;, \quad\quad i,j=1,\ldots,N~.
\end{equation}
The model consists of an ensemble of Hamiltonians 
\begin{equation}
\label{SYK Hamiltonian}
    H_{q} = (i)^{\frac{q}{2}} \sum_{1\leq i_1 < i_2 <\ldots< i_q\leq N} J_{i_1i_2\ldots i_q}\psi_{i_1}\psi_{i_2}\ldots \psi_{i_q}~, \quad\quad q \in 2 \mathbb{Z}^+ ~,
\end{equation}
where the coupling constants $J_{i_1i_2\ldots i_q}$ of the theory are independent and identically distributed random variables drawn from a Gaussian distribution with 
\begin{equation}
\label{coupling distribution}
    \langle J_{i_1i_2\cdot\cdot\cdot i_q}\rangle = 0~, \quad\quad \langle J^{2}_{i_1i_2\cdot\cdot\cdot i_q}\rangle = \frac{2^{q-1}}{q}\frac{\mathcal{J}^2 (q-1)!}{N^{q-1}}~.
\end{equation}
At finite $N$, the dimensionality of the Hilbert space is $2^{N/2}$, which makes it computationally hard to exactly diagonalise the Hamiltonian for $N \gtrsim 36$. In this regime, it can already be observed that the spectrum of the theory is chaotic for all values of $q \geq 4$, and integrable for $q=2$ \cite{Maldacena:2016hyu, Sarosi:2017ykf}. 

In the large-$N$ limit, the theory can be solved in the saddle-point approximation. The partition function can then be expressed in terms of bi-local fields in Euclidean time, $G(\tau_1,\tau_2)$ and $\Sigma(\tau_1,\tau_2)$ \cite{Sachdev:2015efa, Maldacena:2016hyu,Kitaev:2017awl}. These represent the fermion bilinear (defined below) and self energy, respectively. The Euclidean time coordinate $\tau \sim \tau + \beta$ is periodically identified with period given by the inverse temperature $\beta$. Physically, $G(\tau_1,\tau_2)$ computes the (time-ordered) thermal two-point function\footnote{In the large $N$ limit, many quantities are self averaging including the thermal two-point function \cite{Chowdhury:2021qpy}. It is an open question whether the Krylov complexity is self-averaging too. We will be working under this assumption since we use the averaged two-point function as a starting point for our computation of the Krylov complexity.}  
\begin{equation}
\label{SYK autocorrelation}
G(\tau_1,\tau_2) = \frac{1}{N}  \sum_{i=1}^{N} \langle T\psi_i(\tau_1)\psi_i(\tau_2) \rangle_{\beta} \,, 
\end{equation}
averaged over disorder. At large $N$, $G(\tau_1,\tau_2)$ and $\Sigma(\tau_1,\tau_2)$ obey the Schwinger-Dyson equations
    \begin{eqnarray}\label{field equations large N i}
    \begin{cases}
        G^{-1}(\tau_1,\tau_2) =  \delta(\tau_1-\tau_2)\partial_{\tau_2}-\Sigma(\tau_1,\tau_2) ~, \\
       \Sigma(\tau_1,\tau_2) =\mathcal{J}^2 \frac{2^{q-1}}{q}  G(\tau_1,\tau_2)^{q-1}~.
        \end{cases}
    \end{eqnarray}
The above equations can be solved numerically using a recursive algorithm and the fast Fourier transform \cite{Maldacena:2016hyu} or the Legendre spectral method \cite{Cruz:2022uic}. At infinite coupling (or low temperatures), $\beta \mathcal{J} \to \infty$, the Schwinger-Dyson equations develop an extra emergent reparametrisation symmetry, from which it can be seen that $G(\tau_1,\tau_2)$ behaves as a conformal two-point function with conformal weight $\Delta=1/q$.

The SYK model admits further computational control if, after taking the large $N$ limit, we also take the large-$q$ limit.\footnote{Another solvable limit of the SYK model is the double-scaling limit, in which both $N$ and $q$ are taken to infinity but $q^2/N$ is kept finite and fixed \cite{Berkooz:2018qkz,Berkooz:2018jqr}.} In this case, we can expand $G(\tau_1,\tau_2) = G(\tau_1-\tau_2)$ as
\begin{equation}
\label{large q 1st order}
    G(\tau) = \frac{\mbox{sgn}(\tau)}{2}\left(1 + \frac{g(\tau)}{q}+O(1/q^{2})\right)~.
\end{equation}
To leading non-trivial order in $q$, the Schwinger-Dyson equations (\ref{field equations large N i})  become a single ordinary differential equation for $g(\tau)$, namely
\begin{equation} \label{diff_eqn}
    \partial_{\tau}^{2}g(\tau) =  2 \mathcal{J}^2  e^{g(\tau)} \,. 
\end{equation}
Imposing thermal boundary conditions, $g(0)=g(\beta)=0$, we find that, 
\begin{equation}  \label{1st correction large q}
e^{g(\tau)} = \frac{\cos^2 \nu}{\cos^2 \left( 2 \nu \left( \frac{1}{2} - \frac{|\tau|}{\beta} \right)\right)}\, , \quad\quad \beta \mathcal{J} = \frac{2 \nu}{\cos \nu} \,.
\end{equation}
At low temperatures, the parameter $\nu$ can be expanded as follows,
\begin{equation} \label{eq: nu_single}
    \nu=\frac{\pi}{2}-\frac{\pi}{\beta \mathcal{J}}+\frac{2 \pi}{(\beta \mathcal{J})^2}-\frac{\pi\left(24+\pi^2\right)}{6(\beta \mathcal{J})^3}+O(\beta \mathcal{J})^{-4} .
\end{equation}
while at large temperatures $\nu\sim \frac{\beta \mathcal{J}}{2}-\frac{(\beta \mathcal{J})^3}{16}+\frac{13 (\beta \mathcal{J})^5}{768}+  \cdots$. This can be used to extract the low-temperature thermodynamic behaviour of the theory. In particular, the thermal entropy at low temperatures reads \cite{Anninos:2022qgy}
\begin{equation}
\label{entropy large q}
    \frac{S_{\text{thermal}}}{N} = \left(\frac{\log 2}{2} -\frac{\pi^2}{4 q^2}\right) + \frac{\pi^2}{q^2}\frac{1}{\beta \mathcal{J}} + \cdot\cdot\cdot ~.
\end{equation}
It is also possible to include higher order corrections in $1/q$ \cite{Tarnopolsky:2018env}.  
At finite $q$, qualitatively the same behaviour for the entropy holds. Namely, there is a non-vanishing entropy at zero temperature, followed by a correction that is proportional to $(\beta \mathcal{J})^{-1}$. The exact coefficients can be found numerically for different values of $q \geq 4$ \cite{Maldacena:2016hyu,Tarnopolsky:2018env}.
One of the salient features of the SYK model is that there is enough computational control to compute dynamical quantities such as the OTOC at large $N$ and in the strongly-coupled regime. 
At infinitely low temperatures, where conformal symmetry is emergent, it can be checked that the SYK model saturates the maximal chaos bound \cite{Maldacena:2015waa}. The leading order correction away from maximal chaos can also be computed analytically and it gives, \eg for $q=4$, $q=6$, and $q\to \infty$ \cite{Maldacena:2016hyu}, 
\begin{equation}
\label{lambda_C}
\lambda_C\equiv
\left(\lambda_{L}\right)_{\beta \mathcal{J}\gg 1} \approx \begin{cases}\frac{2\pi}{\beta} \left( 1-\frac{4.28}{\beta\mathcal{J}}+\dots \right) \,,\quad q=4~,\\
\frac{2\pi}{\beta} \left( 1-\frac{3.11}{\beta\mathcal{J}}+\dots \right) \,,\quad q=6~. \\
\frac{2\pi}{\beta} \left( 1-\frac{2}{\beta\mathcal{J}}+\dots \right) \,,\quad q\to \infty~.
\end{cases}
\end{equation}
Away from this limit, the Lyapunov exponent can be computed numerically, see appendix \ref{sec: comp Lyapunov}. In the large-$q$ limit, it is possible to compute it analytically for all temperatures, obtaining $\lambda_L = \frac{4\nu}{\beta}$.  

\subsubsection{Deformed SYK models} \label{subsec:def_SYK}

A \textit{relevant} question is whether the SYK model admits deformations away from its near-conformal infrared fixed point. Simple operators in the SYK model are mostly irrelevant \cite{Rosenhaus:2018dtp}. However, it is possible to deform the SYK model with deformations that take the form of the SYK Hamiltonian itself, but with a different number of fermions. Namely, it is possible to consider the following theory,
\begin{equation}
\label{deformed Hamiltonian}
    H_{\mathrm{def}}=H_{q}+s H_{\tilde{q}}  ~,
\end{equation}
where $s$ is a tunable dimensionless parameter and the Hamiltonian $H_x$ denotes the Hamiltonian (independent random) ensemble \eqref{SYK Hamiltonian}-\eqref{coupling distribution} of a single SYK model with $x$-fermion interactions.\footnote{Similar deformations have been recently considered in the double-scaling limit \cite{Berkooz:2024evs,Berkooz:2024ofm}.} Note that the Hamiltonian is built from the same $N$ fermions.

In this paper, we will assume that $q > \tilde{q}$. A naive power counting argument indicates that this is required for the second term to be relevant. The argument goes as follows. Near the conformal fixed point of the first SYK, the fermions acquire a scaling dimension of $\Delta=1/q$, so the second SYK Hamiltonian, thought of as a disordered operator \cite{Anninos:2020cwo,Anninos:2022qgy}, has a naive scaling dimension of $\Delta = \tilde{q}/q < 1$, and will become dominant in the infrared. This intuition was proven to be correct, at least for small values of $s$, both at large and finite values of $q$ \cite{Anninos:2020cwo,Anninos:2022qgy}. This hints that for sufficiently large $q$, the SYK model could admit relevant deformations that consist of a sum of multiple SYK Hamiltonians. These deformations could be  tractable through the different regimes of the RG flow. We will come back to this in section \ref{sec:outlook}.

For now, let us restrict the analysis to a single deformation. In this case, it is also possible to study the theory in the large-$N$ limit, where the deformed Schwinger-Dyson equations become
\begin{eqnarray}
    \begin{cases}
           G^{-1}(\tau_1,\tau_2) =  \delta(\tau_1-\tau_2)\partial_{\tau_2}-\Sigma(\tau_1,\tau_2) ~, \\
       \Sigma(\tau_1,\tau_2) =\mathcal{J}^2 \left(\frac{2^{q-1}}{q}  G(\tau_1,\tau_2)^{q-1}  +  s^{2} \frac{2^{\tilde{q}-1}}{\tilde{q}} G(\tau_1,\tau_2)^{\tilde{q}-1}\right) ~.
    \end{cases} \label{SD2}
\end{eqnarray}
A case of special interest is  $\tilde{q}=2$. The deformation then consists of an integrable Hamiltonian, often referred to as mass-deformed SYK, and has been broadly studied in the context of quantum chaotic to integrable transitions \cite{Garcia-Garcia:2017bkg, Kim:2020mho, Garcia-Garcia:2020dzm, Lunkin:2020tbq, Nandy:2022hcm,Menzler:2024atb}. 

The deformed SYK model can also be studied in the large-$q,\tilde{q}$ limit \cite{Jiang:2019pam,Anninos:2020cwo, Anninos:2022qgy,Louw:2023lpq}. It is convenient to define the ratio $\mathfrak{n} \equiv q/\tilde{q}>1$ and take both $q$ and $\tilde{q}$ to infinity, while keeping $\mathfrak{n}$ fixed. In this case, the Schwinger-Dyson equations \eqref{SD2} reduce to
\begin{equation} \label{deformed_diff_eqn}
    \partial_{\tau}^{2}g(\tau) =  2\mathcal{J}^2 \left( e^{g(\tau)} + \mathfrak{n} s^2 e^{g(\tau)/\mathfrak{n}}\right)~. 
\end{equation}
In general, this differential equation (supplemented with thermal boundary conditions) can be solved numerically using standard shooting methods \cite{Anninos:2022qgy}. Remarkably, there is at least one case that can be solved analytically for all values of $\beta \mathcal{J}$ and $s$. This is when $\mathfrak{n}=2$, which means that the number of fermions in the relevant deformation is half of that in the original Hamiltonian  \cite{Jiang:2019pam, Anninos:2020cwo}. In this case, the two-point function is given by 
\begin{equation}\label{largeq1}
e^{g(\tau)}=\frac{4 \nu ^4}{\left(\sqrt{(\beta\mathcal{J})^2\nu^2+s^4(\beta\mathcal{J})^4}\cos\left(\nu\left(\frac{2|\tau|}{\beta}-1\right)\right)+s^2(\beta\mathcal{J})^2\right)^2} \,,
\end{equation}
where the parameter $\nu$ is defined as
\begin{equation}\label{largeqcon}
\cos \nu= \frac{2 \nu^2 - s^2(\beta\mathcal{J})^2}{\sqrt{(\beta\mathcal{J})^2\nu^2+s^4(\beta\mathcal{J})^4}}.
\end{equation}
Paralleling the analysis of the single SYK model, we can compute the thermodynamic behaviour of this model at low temperatures. In fact, provided that $s \ll 1$, there are two regimes in which the theory is nearly conformal and the entropy becomes (nearly) linear in the temperature. We refer to these two regimes as the \textbf{intermediate} ($1/s^2\gg\beta J\gg 1$) and the \textbf{deep} infrared ($\beta J\gg 1/s^2$) of the theory.

To extract the intermediate infrared behaviour analytically, we expand the parameter $\nu$ above for small values of $s$ and large values of $\beta \mathcal{J}$ such that $\beta \mathcal{J} s$ is kept fixed (of order one). This expansion yields
\begin{align} \label{nu_int}
    \nu_{\text{Int IR}}=&\frac{\pi }{2}+\frac{2 (\beta \mathcal{J} s)^2-\pi ^2}{\pi  \beta  \mathcal{J}}+\frac{2}{\pi ^3} \frac{\left(\pi ^4-4 (\beta \mathcal{J} s)^4\right)}{(\beta \mathcal{J})^2} \nonumber
    \\&+
    \frac{16 \left(24-\pi ^2\right) (\beta  \mathcal{J} s)^6-96 \pi ^2 (\beta   \mathcal{J} s)^4+6 \pi ^6 (\beta \mathcal{J} s)^2-\pi ^6 \left(24+\pi ^2\right)}{6 \pi ^5 (\beta  \mathcal{J})^3}+\ldots
\end{align}
In this case we obtain for the entropy \cite{Anninos:2020cwo}
\begin{equation}
    \text{Intermediate IR:} \,\,\, \frac{S_\text{thermal}}{N}=\left(\frac{\log 2}{2}-\frac{\pi^{2}}{4q^2}\right) +\frac{\pi^{2}}{q^2}\frac{1}{\beta \mathcal{J}}  -\frac{2s^2 \beta \mathcal{J}}{q^2}+\cdots \,,
\end{equation}
which has the linear-in-temperature behaviour expected for a single SYK theory with $H_q$, and $s$ will only appear as a small correction away from this intermediate near fixed point. For lower temperatures, the theory develops a new near fixed point, that can be studied analytically by doing the following expansion,
\begin{equation} \label{nu_deep}
 \nu_{\text{Deep IR}}=\pi-\frac{\pi \sqrt{1+4 s^2}}{s^2 \beta \mathcal{J}}+\frac{\pi(4  s^2+1)}{s^4} \frac{1}{(\beta \mathcal{J})^2}+\frac{\pi ^3 \left(-2 s^4+2 s^2+1\right)-3 \pi  \left(4 s^2+1\right)^2}{3 s^6 \sqrt{4 s^2+1}} 
\frac{1}{(\beta \mathcal{J})^3}+ \cdots ~,
\end{equation}
in this case the entropy becomes
\begin{equation}
\label{deformed deep IR entropy}
  \text {Deep IR:} \,\,\, \frac{S_\text{thermal}}{N} = \left(\frac{\log 2}{2} -\frac{\pi^{2}}{4\tilde{q}^{2}}\right)+ \frac{\sqrt{1+4s^2}}{2s^2}\frac{\pi^{2}}{\tilde{q}^{2}} \frac{1}{\beta \mathcal{J}}+\cdots  \, .
\end{equation}
The full behaviour of the thermal entropy as a function of both couplings can be found in \cite{Anninos:2020cwo}. As with the single SYK, in this model it is also possible to compute dynamical quantities associated with quantum chaos. In particular, the Lyapunov exponent can be computed analytically around both near fixed points \cite{Jiang:2019pam}, 
\begin{equation} \label{lyapunov_def}
    \lambda_L = \begin{cases} \frac{2\pi}{\beta} \left( 1 - \frac{2}{\beta\mathcal{J}} - \left(\frac{1}{2}-\frac{4}{\pi^2}\right)s^2\beta\mathcal{J} + \cdots \right)
 \,, & \text{Intermediate IR} \,, \\
 \frac{2\pi}{\beta} \left( 1 - \frac{1}{s^2\beta\mathcal{J}} + \cdots\right)
 \,, & \text{Deep IR} \,.
\end{cases}
\end{equation}
This shows, that close enough to both fixed points the system becomes maximally chaotic, while in between the Lyapunov exponent decreases to non-maximal values. Away from these limits, the full behaviour of the Lyapunov exponent was also found numerically in \cite{Jiang:2019pam}, which we reproduce in section \ref{sec:Chaos-to-chaos flows}  and compare to the Krylov computations.

\subsection{Review of Krylov complexity}

In this section we review the notion of Krylov complexity and its relation to chaos in quantum systems. We will be interested in the growth of a simple operator $\mathcal{O}$ under Heisenberg time evolution. Heisenberg time evolution spreads the operator through a series of nested commutators, which can be written in terms of the Liouvillian operator $\mathcal{L}:=[H,\cdot]$ as
\begin{equation}
\label{nested basis}
\mathcal{O}(t) = e^{iHt}\mathcal{O}e^{-iHt} = \sum_{n=0}^{\infty}\frac{(it)^n}{n!}\mathcal{L}^{n}\mathcal{O}~\,.
\end{equation}
In the following we will always assume the Hamiltonian and the operator $\mathcal{O}$ to be Hermitian.
To describe the operator spreading, we first consider the vector space spanned by the nested commutators
\begin{equation}
\mathcal{H}_{\mathcal{O}}=\text{span}\{\mathcal{O},[H, \mathcal{O}],[H,[H, \mathcal{O}]], \ldots\}~.
\end{equation}
This operator space is called  the Krylov subspace and it contains the time-evolved operator $\mathcal{O}(t)$ for all $t$. In order to quantify the operator growth within this subspace one defines an inner product acting on operators in the theory. The inner product used at infinite temperature is given by 
\begin{equation}
\label{inner product infinite temp}
(\mathcal{O}_1|\mathcal{O}_2):=\frac{\text{Tr}(\mathcal{O}_1^{\dagger}\mathcal{O}_2)}{\text{Tr}(I)},
\end{equation}
where $I$ is the identity operator. One can also consider the system at finite inverse temperature $\beta$, for which it is usually useful to use the Wightman inner product,
\begin{equation}
\label{wightman inner product}
\left(\mathcal{O}_1|\mathcal{O}_2 \right)_{\beta}^{W} := \frac{1}{Z}\text{Tr}\left(e^{-\beta H/2}\mathcal{O}_1^{\dagger}e^{-\beta H/2}\mathcal{O}_2\right)~,
\end{equation} 
where $Z = \text{Tr} \left(e^{-\beta H}\right)$.
Taking the $\beta\to0$ limit recovers the inner product at infinite temperature. One can then find an orthonormal basis with respect to the inner product via the Gram-Schmidt procedure. We will refer to this basis as the Krylov basis. For a Hermitian operator, the Gram-Schmidt procedure applied to \eqref{nested basis} is described in \cite{Parker:2018yvk} as follows. First define
\begin{eqnarray}
\left.|\mathcal{O}_0\right):=  \left.|\mathcal{O}\right),\; b_{1} = \left(\mathcal{O}_0 \mathcal{L}|\mathcal{L}\mathcal{O}_0\right),\; \left.|\mathcal{O}_1\right):=b_{1}^{-1}\mathcal{L}\left.|\mathcal{O}_0\right) \,.
\end{eqnarray}
Then, for $n>1$, carry out the following recursive algorithm 
\begin{equation}
\label{krylov gs}
\begin{cases}
\left.| A_n\right) & \left.\left.:=\mathcal{L} | \mathcal{O}_{n-1}\right)-b_{n-1} | \mathcal{O}_{n-2}\right) \,, \\
b_n & :=\left(A_n | A_n\right)^{1 / 2} \,, \\
\left.| \mathcal{O}_n\right) & \left.:=b_n^{-1} | A_n\right)~.
\end{cases}
\end{equation}
For finite dimensional Hilbert spaces, the recursive algorithm stops for some $n=D_K$ known as the Krylov dimension.

The $b_n$'s are called Lanczos coefficients and from \eqref{krylov gs} it is easy to see that the Liouvillian operator is tridiagonal with respect to the Krylov basis with the upper and lower diagonal entries corresponding to the Lanczos coefficients,
\begin{equation}\label{momentsmat}
\left(\mathcal{O}_n|\mathcal{L}| \mathcal{O}_m\right)=\left(\begin{array}{ccccc}
0 & b_1 & 0 & 0 & \cdots \\
b_1 & 0 & b_2 & 0 & \cdots \\
0 & b_2 & 0 & b_3 & \cdots \\
0 & 0 & b_3 & 0 & \ddots \\
\vdots & \vdots & \vdots & \ddots & \ddots
\end{array}\right)~.
\end{equation}
The time-dependent operator admits the following expansion, 
\begin{align}
	\mathcal O(t)=\sum_{n=0}^{D_K-1} i^n\varphi_n(t)|\mathcal O_n)\,, 
\end{align}
where all the time-dependence has been completely transferred to ``wavefunctions'' $\varphi_n(t)$ of the Schr\"odinger equation of an effective one dimensional tight-binding chain
\begin{align}\label{eq:tightbindingchain}
	\partial_t \varphi_n(t)=b_n\varphi_{n-1}(t)-b_{n+1}\varphi_{n+1}(t)\,, \qquad \varphi_n(0) = \delta_{n0}\,, \quad \varphi_{-1}(t)=0.
\end{align}
From this one can directly infer that the spread of the operator in time over the Krylov-basis is determined by the Lanczos coefficients. The faster the Lanczos coefficients grow with $n$, the faster the operator will spread over the Krylov basis. 

In \cite{Parker:2018yvk}, it was noted that in non-integrable many-body systems 
the Lanczos coefficients are asymptotically linear in $n$,\footnote{Note that for one-dimensional systems, there could be a logarithmic correction to the linear growth \cite{Nandy:2024htc}. This correction is not present in SYK-like models.} 
\begin{equation}
\label{operator growth hypothesis}
b_n \sim \alpha \, n + \gamma ~,
\end{equation}
where the slope $\alpha>0$ and $\gamma$ are constant real numbers. 
When the system is integrable, the Lanczos coefficients will grow with a fractional power $b_n\sim n^\delta$, for $0<\delta<1$. A linear growth in the Lanczos coefficient\footnote{This assumes that the dependence of the Lanczos coefficients on $n$ is smooth.} leads to an exponential growth in the Krylov complexity that is defined as,
\begin{align}\label{eq: def CK}
	C_K\equiv \sum_{n=0}^{D_K-1} n|\varphi_n|^2\;\,\stackrel{b_n\sim \alpha \, n}{\sim}\;\,e^{\lambda_Kt}\,,
\end{align}
where we defined the Krylov exponent as follows\footnote{One should note that there are known examples where the exponential of the Krylov complexity $\lambda_K$ does not satisfy this relation with the slope of the Lanczos coefficients $\alpha$. This was shown to happen in thermal quantum field theory, where the Lanczos coefficients split into separate branches for even and an odd $n$ \cite{Avdoshkin:2022xuw,Camargo:2022rnt}, where the staggering was proposed to be related to the mass gap in the power spectrum. 
We will indeed see such a splitting for the finite $q$ SYK, but both even and odd coefficients will be linear with the same slope and the staggering effect will be very small in all the cases we study in this paper. Therefore, we will assume that $\lambda_K = 2\alpha$. We leave a full evaluation of $\lambda_K$ for future work.\label{footy}}
\begin{equation}
    \lambda_K \equiv 2\alpha\, .
\end{equation}
The Krylov complexity $C_K$ measures the average position of the operator in the tight-binding chain, which effectively describes the spread of the operator over the Krylov basis. 
Note that, as a consequence, in integrable quantum systems, the Krylov complexity typically does not grow exponentially. See \cite{Nandy:2024htc} for different known behaviours. 

For chaotic systems where \eqref{operator growth hypothesis} holds (and as long as the Lyapunov exponent is well-defined), it was proposed that the Krylov exponent could provide an upper bound for the Lyapunov exponent, see \eqref{bound} that we recall here,
\begin{equation}
\mathcal{\lambda}_{L}\leq \lambda_K \leq \frac{2\pi}{\beta} ~.
\end{equation}
The left inequality  was proven at infinite temperature and conjectured to hold at finite temperature in \cite{Parker:2018yvk}. Under the assumption of certain analytic and smoothness properties of the Lanczos coefficients $n$, the right inequality was later proven at finite temperature in \cite{Avdoshkin:2019trj,Gu:2021xaj}.

In this paper, we will compute $\lambda_K$ in different quantum systems where $\lambda_L$ can also be computed, to gain insight into how tight is the bound, especially in systems where the behaviour of $\lambda_L$ is not necessarily monotonic, and could even go to zero. Before doing so, we review two methods to compute $\lambda_K$, one based on the explicit calculation of the Lanczos coefficients and the other based on analytic properties of the Wightman two-point function.

\subsubsection{Krylov exponent from moments}
\label{sec: lanczos from moments}

Performing the algorithm \eqref{krylov gs} can be  numerically challenging. Instead, if we know the Wightman autocorrelation function, $C(t)$, defined by
\begin{equation}
C(t)=\left(\mathcal{O}(t)|\mathcal{O}(0)\right)^{W}_{\beta}~,
\end{equation}
the Lanczos coefficients can be computed using a recursive algorithm \cite{Viswanath:1994}. The first step is to analytically continue to imaginary time and perform a Taylor expansion to find the moments, $\mu_{2n}$,
\begin{equation}
\label{taylor expansion}
C(-i \tau)=\sum_{n=0}^{\infty} \mu_{2 n} \frac{\tau^{2 n}}{(2 n) !}~,
\end{equation}
where the odd coefficients vanish since we have assumed that $\mathcal{O}$ is Hermitian. 
The moments are therefore related to expectation values of powers of the Liouvillian 
$\mu_{2n}:=(\mathcal{O} |\mathcal{L}^{2n}|\mathcal{O} ) =\frac{d^{2n}}{d\tau^{2n}}C(-i\tau)|_{\tau=0}$
and so products of the Lanczos coefficients are given by determinants of minors of the Hankel matrix of moments: 
$b_1^{2n}....b_n^2=\det(\mu_{i+j})_{0 \leq  i,j \leq n}$.\footnote{In this expression we fixed a small  typo in equation (A4) of \cite{Parker:2018yvk}, see \eg (2.10) of \cite{Caputa:2024vrn}.} 
Therefore, after obtaining the moments the following recursive algorithm can be used to obtain the Lanczos coefficients, $b_n$:
\begin{equation}
\label{coeffs from moments}
\begin{cases}
\text{Goal}: \quad  b_n = \sqrt{M_{2n}^{(n)}},\\
\text{Recursive step}: \quad M_{2k}^{(n)} = \frac{M_{2k}^{(n-1)}}{b^{2}_{n-1}} - \frac{M_{2k-2}^{(n-2)}}{b^{2}_{n-2}},\\
\text{Stopping conditions}:  \quad M_{2k}^{0} = \mu_{2k},\; M_{2k}^{(-1)} = 0,\; b_{-1}=b_0 = 1.
\end{cases}
\end{equation}
Once sufficiently many Lanczos coefficients are obtained in this way, the Krylov exponent, $\lambda_{K}$ can be found by computing the slope of the coefficients, see however footnote \ref{footy}. 

In what follows we will study the Krylov complexity for the SYK model and its deformations with respect to a single fermion operator $\mathcal{O} = \sqrt{2} \psi_1$. In the large $N$, limit the autocorrelation function 
$\langle \psi_i(\tau)\psi_i(0)\rangle$ is independent of $i$ \cite{Chowdhury:2021qpy} and therefore
\begin{equation}
\label{autocorrelation}
 C(-i\tau) = 2G(\tau + \beta/2)~,
\end{equation}
where $G(\tau)\equiv G(\tau,0)$ was defined in \eqref{SYK autocorrelation}. 
The moments used as a starting point for the algorithm \eqref{coeffs from moments} are computed as follows
\begin{equation}\label{momentdef}
\mu_{2n}=\frac{d^{2n}}{d{\tau}^{2n}}\left(2G(\tau + \beta/2)\right)|_{\tau=0}~.
\end{equation}

\subsubsection{Krylov exponent from pole of autocorrelation function}\label{sec: slope from pole}
If the Lanczos coefficients $b_n$ have a linear asymptotic growth of the form \eqref{operator growth hypothesis}, then the asymptotic slope $\alpha$ can be extracted from the location of the first pole of the autocorrelation function in Euclidean time \cite{Parker:2018yvk,Dymarsky:2021bjq,Avdoshkin:2019trj,Avdoshkin:2022xuw}. This can be seen by relating the moments $\mu_{2n}$ to the Lanczos coefficients $b_n$. Assuming \eqref{operator growth hypothesis} the moments have the following asymptotic form \cite{Parker:2018yvk},
\begin{equation}
\label{asymopttotic moments}
\mu_{2 n}=\left(\frac{4 n \alpha}{e \pi}\right)^{2 n} e^{\mathrm{o}(n)}~.
\end{equation}
Using \eqref{asymopttotic moments}, one can then apply the root test to \eqref{taylor expansion} to see that the radius of convergence and hence the location of the first pole, which we denote by $\tau_{*}$, is given by 
\begin{equation}
\tau_{*} = \frac{\pi}{2\alpha}~.
\end{equation}
From this one can compute the Krylov exponent using the relation 
\begin{equation}\label{SDrel}
   \lambda_K=\frac{\pi}{\tau_*} \,.
\end{equation}
For the SYK and its deformations, from \eqref{autocorrelation} we see that $\tau_*$ is the first pole of $G(\tau+\beta/2)$. See \cite{Bhattacharjee:2022vlt} for subtleties arising in systems with saddle-dominated scrambling.

\section{Krylov complexity of a single SYK model}\label{sec:singleSYK}
\subsection{The integrable $q=2$ model}\label{sec:The_integrable_q2_model}
\begin{figure}[H]
        \centering
                \includegraphics[scale=0.6]{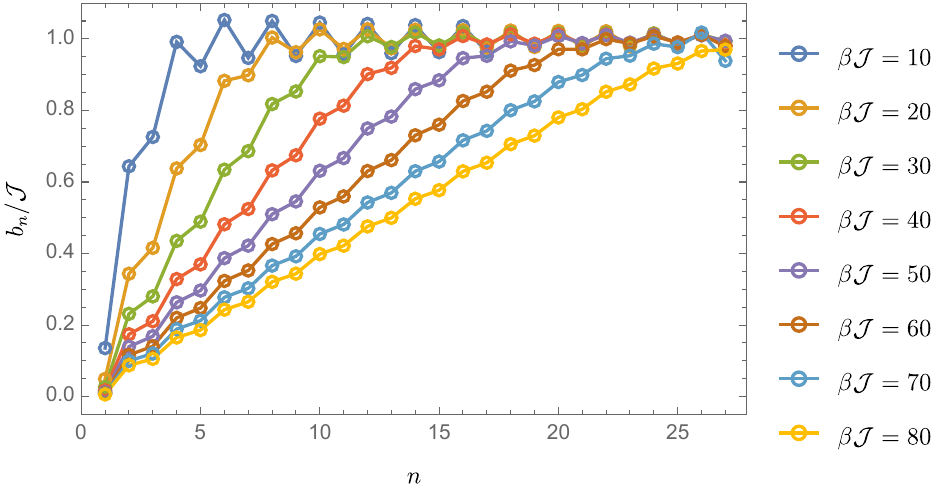}              
                \caption{Lanczos coefficients $b_n$ for the single SYK with $q=2$ for different values of $\beta \mathcal{J}$. }\label{fig:q2_lanczos}   
\end{figure}

\begin{figure}[H]
    \centering
    \includegraphics[scale=0.5]{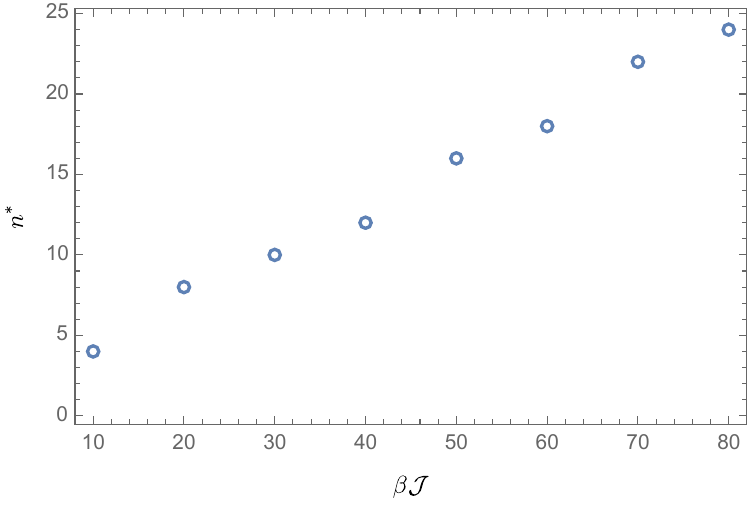}
    \caption{The saturation point $n^*$ as a function of $\beta \mathcal{J}$. The saturation point is defined as the first $n$ such that $|b_n/\mathcal{J} -1| \leq 0.1$.}
    \label{fig:finite_q_2_saturation_pts}
\end{figure}

Consider the single SYK Hamiltonian \eqref{SYK Hamiltonian}. When $q=2$, the interaction term becomes just a random mass term for the fermions, and the model is known to be integrable. We thus expect the Lanczos coefficients to grow sub-linearly with $n$, for large enough $n$. At infinite temperature, this was already demonstrated in \cite{Parker:2018yvk}. Here we consider the theory at finite inverse temperature $\beta$. For $q=2$, the thermal two-point function is known analytically \cite{Maldacena:2016hyu},
  \begin{align} \label{q=2_two_point}
  	G(\tau)=\int_0^\pi \frac{\mathrm d\theta}{\pi}\cos^2\theta\frac{\cosh[(\tfrac{\tau}{\beta}-\tfrac{1}{2})2\beta \mathcal{J} \sin \theta]}{\cosh(\beta \mathcal{J} \sin\theta)}\,.
  \end{align}
We can use the two methods described in sections \ref{sec: lanczos from moments} and \ref{sec: slope from pole} to compute the Krylov exponent. 
Using the first method, we computed the first $\sim 25$ Lanczos coefficients, for a range of different temperatures. The results are shown in Fig.~\ref{fig:q2_lanczos}. We see that $b_n/\mathcal{J}$ initially grows with $n$ but then reaches a plateau around the value of 1. As we increase $\beta\mathcal{J}$ the value of $n$ at which the plateau is reached increases approximately linearly with $\beta \mathcal{J}$, as can be seen from Fig.~\ref{fig:finite_q_2_saturation_pts}.

This means that at sufficiently low temperatures, the number of Lanczos coefficients required to see the saturation can go beyond computational control. One might be fooled to think that the system becomes chaotic at low temperatures, since the available Lanczos coefficients seem to behave linearly with $n$. However, this is not the case. To verify this, it is useful to use the second method, which directly gives the asymptotic behaviour of the slope of the Lanczos coefficients. Since we have the analytic form of the autocorrelation function this is not hard to compute. We need to find the location of the pole in Euclidean time that is closest to the origin in \eqref{q=2_two_point}. 

At any temperature, the only divergence of the two-point function is at $\tau_* \to \infty$. From \eqref{SDrel}, this trivially gives that the slope of the Lanczos coefficients is zero and so, as expected, $\lambda_K = 0$ at any temperature for the single SYK with $q=2$.

This behaviour of the Lanczos coefficients provides a sharp distinction from that of chaotic many-body systems, for which the Lanczos coefficients are expected to grow indefinitely according to the operator growth hypothesis \eqref{operator growth hypothesis}. 

Finally, let us remark that in \cite{Jian:2020qpp}, it was found that at \textit{finite} but large $N$, with $q=4$, the Lanczos coefficients do in fact reach a plateau. In this case however, this is not due to the system being integrable but rather that the Hilbert space is finite. 
In particular they show that the value of this plateau grows linearly with $N$ suggesting that they  would not get saturation in the large $N$ limit.

\subsection{Large $q$ and $1/q$ corrections}\label{sec:single_large_q_corr}

Now we turn to the discussion of chaotic SYK models. The first example analysed in the context of Krylov complexity was the large-$q$ model, as it is possible to get analytic expressions for the autocorrelation function, see \eqref{1st correction large q}.

It was shown that both at finite and infinite temperature, the resulting Lanczos coefficients display the expected linear growth \cite{Parker:2018yvk}. Using the moments method, the  Lanczos coefficients at infinite temperature  take the form
\begin{equation}
\label{large q lanczos infinite temp}
b_n/\mathcal{J}= \begin{cases} \sqrt{2 / q}+O(1 / q)\,, & n=1 \,, \\ \sqrt{n(n-1)}+O(1 / q)\,, & n>1 \,. \end{cases}
\end{equation}
At finite temperature, we have that 
\begin{equation}
\label{large q lanczos finite temp}
b_n/\mathcal{J}= \begin{cases}\frac{2 \nu}{\beta\mathcal{J}} \sqrt{2 / q}+O(1 / q)\,, & n=1 \,, \\ \frac{2 \nu}{\beta\mathcal{J}} \sqrt{n(n-1)}+O(1 / q)\,, & n>1\,.\end{cases}
\end{equation}
By considering the asymptotic form of the coefficients for large $n$, it follows that 
\begin{equation}
\label{large_q_krylov}
\lambda_{K} = \frac{4\nu}{\beta} \,.
\end{equation}
This value for the Krylov exponent can also be verified by looking at the pole of the 
autocorrelation function \eqref{1st correction large q}, with the Euclidean time shifted to compute the Wightman correlator, see \eqref{autocorrelation}. In fact, one finds that
\begin{equation}
    \tau_* = \frac{\beta \pi}{4\nu} \longrightarrow \lambda_K = \frac{4\nu}{\beta} \,.
\end{equation}
It is easy to find the low-temperature behaviour of $\lambda_K$ by using the expansion of $\nu$ in \eqref{eq: nu_single},
\begin{equation} \label{single_SYK_krylov_expansion}
    \lambda_K = \frac{2\pi}{\beta} \left(1 - \frac{2}{\beta \mathcal{J}} + \frac{4}{(\beta \mathcal{J})^2} - \frac{24+\pi^2}{3(\beta \mathcal{J})^3} + \cdots \right) \,.
\end{equation}
We see that the Krylov exponent, in the strict large-$q$ limit, provides a far better bound for the Lyapunov exponent than the chaos bound \eqref{BCHAOS}. Indeed,  the Krylov exponent saturates the left inequality in \eqref{bound}, 
\begin{equation}\label{tightbound}
    \lambda_L = \lambda_K \leq \frac{2\pi}{\beta} \,,
\end{equation}
yielding an optimal bound for the Lyapunov exponent, while the chaos bound \eqref{BCHAOS} is only reached, in this case, in the zero temperature limit.

\

\noindent \textbf{Corrections in $1/q$.} 
It is natural to ask if the tight bound \eqref{tightbound} holds beyond the large $q$-limit. 
To answer this question, we consider the next-to-leading contributions in $1/q$ to the thermal autocorrelation function, which were computed in \cite{Tarnopolsky:2018env}. We will consider the full finite-$q$ result numerically in the next section.

We expand the autocorrelation function as follows,
 \begin{equation}
 G(\tau) = \frac{1}{2}\left(1+\frac{g(\tau)}{q}+\frac{h(\tau)}{q^2}+O(1/q^3)\right) ~,
 \end{equation}
where the leading order $g(\tau)$ is given in \eqref{1st correction large q} while $h(\tau)$ is found to be \cite{Tarnopolsky:2018env}
 \begin{equation}\label{2nd correction large q}
 \begin{aligned}
 h(\tau)&=\frac{1}{2}g^2(\tau)-2\ell(\tau)-4\left(\tan \left(\nu-\frac{2\nu \tau}{\beta}\right) \int_{\beta /2}^{\tau}  \left(-\frac{2\nu}{\beta}\right)\ell(y)\; dy+1\right)\\
 &+4\frac{1+\left(\nu-\frac{2\nu \tau}{\beta}\right) \tan \left(\nu-\frac{2\nu\tau}{\beta}\right)}{1+\nu \tan \nu}\left(\tan \nu \int_{\beta/2}^{0}\left(-\frac{2\nu}{\beta}\right)\ell(y)\; dy+1\right),
 \end{aligned}
\end{equation}
with $\ell(\tau) \equiv g(\tau)-e^{-g(\tau)}\text{Li}_2(1-e^{g(\tau)})$ and 
\begin{equation}
\int_{\beta/2}^{0}\left(-\frac{2\nu}{\beta}\right)\ell(y)\; dy=-\frac{\nu^2}{6 \cos ^2 \nu}(2\nu+3 \sin 2\nu)\,.
\end{equation}

\noindent
Here $\nu$ is the same as in the large-$q$ SYK, given implicitly as a function of the inverse temperature in equation \eqref{1st correction large q}.

\begin{figure}[H]
        \centering
         \subfigure[]{
                \includegraphics[scale=0.55]{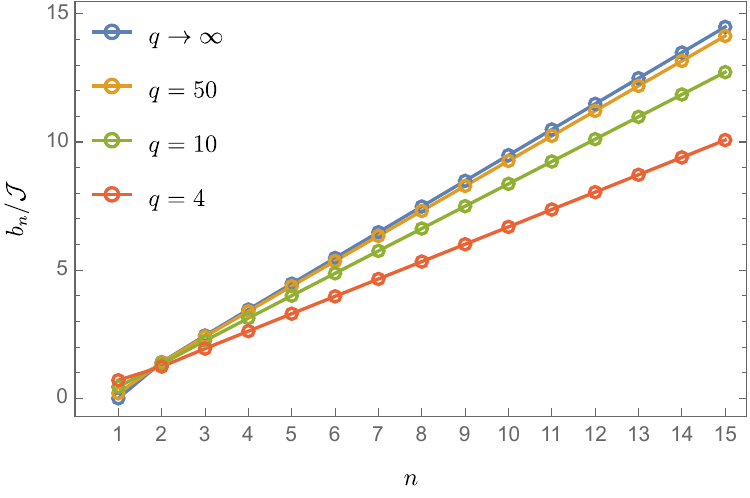}\label{fig:large_q_lanczos_infT(a)}}  \quad\quad
       \subfigure[]{
                \includegraphics[scale=0.55]{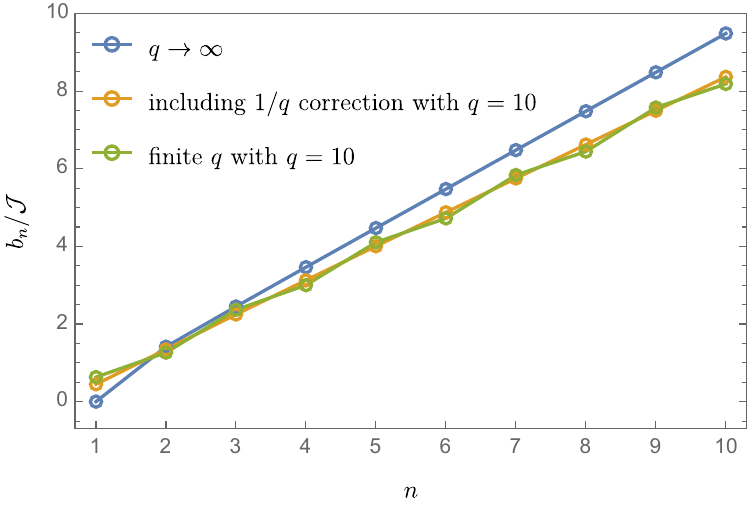} \label{fig:large_q_finite_q_comparison(b)}}   
                
                \caption{Lanczos coefficients as a function of $n$, computed perturbatively in the large-$q$ expansion including the $1/q$ correction at infinite temperature. In (a) we plot different values of $q$ while in (b) we compare to the finite $q=10$ result.}\label{fig:large_q_lanczos_infT}
\end{figure}

As before, we evaluate the moments and Lanczos coefficients from $G(\tau)$, which now includes a correction at order $1/q$. 
Note that when evaluating the moments, the integral $\int_{\beta /2}^{\tau}  \left(-\frac{2\nu}{\beta}\right)\ell(y)\; dy$
does not need to be computed since the moments $\mu_{2n}$ for $n>1$ are extracted by taking $\tau$-derivatives and for the case of $\mu_0$ the integral vanishes at $\tau=\beta/2$. 
Fig.~\ref{fig:large_q_lanczos_infT(a)} shows the first 15 Lanczos coefficients at large $q$ and infinite temperature including the $1/q$ corrections for different values of $q$.

Note that the Lanczos coefficients at infinite temperature and large $q$ (including the next-order correction) differ significantly from those found in \cite{Bhattacharjee:2022ave}. We do not observe a super-linear growth and sub-linear growth in the even and odd Lanczos coefficients, respectively. In fact we find a linear dependence and do not observe any staggering.

\begin{figure}[H]	
    \centering
    \vspace{0mm}
    \includegraphics[width=0.45\columnwidth]{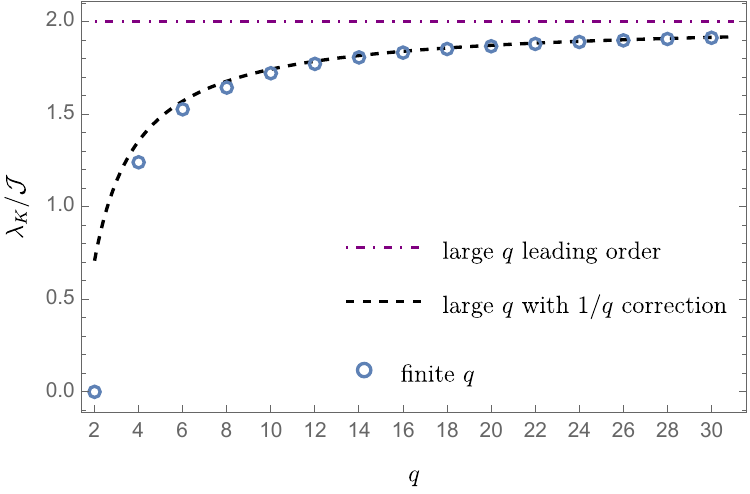} \caption{Comparison of the Krylov exponent at infinite temperature between the large-$q$ result with 1/$q$-correction and finite $q$ results for different values of $q$. The dashed-dotted purple line at $\lambda_K/\mathcal J=2$ denotes the value of the Krylov exponent to leading order in the large-$q$ expansion.}\label{fig:krylov_large_q}
\end{figure}

We can compare our large-$q$ results with numerical results obtained at finite $q$ (see section \ref{sec: Single SYK and saturation of Lyapunov exponent}). For $q=10$ at infinite temperature, the results are shown in Fig.~\ref{fig:large_q_finite_q_comparison(b)}, finding good agreement between the large-$q$ expansion and the finite $q$ numerical result. In Fig.~\ref{fig:krylov_large_q} we compare the large-$q$ Krylov exponent (computed from the slope of the Lanczos sequence) at infinite temperature to the finite $q$ results for different values of $q$, again finding good agreement for $q\gtrsim 10$. Note that in both cases, in order to get this agreement, it was essential to also include the $1/q$-correction. This shows that, at least at infinite temperature, the results obtained in the large-$q$ expansion for the Lanczos coefficients are reasonable and a good approximation to the finite-$q$ results. 

Next, we move away from infinite temperature. In fact, using the expansion in $\nu$, we can obtain analytic results for the first Lanczos coefficients at small $\beta \mathcal{J}$. The first five Lanczos coefficients as $\beta \mathcal{J} \to 0$ are given by, 
\begin{equation}
\begin{cases}
b_1/\mathcal{J} &= \left(\sqrt{2}-\frac{\left(\beta\mathcal{J}\right) ^2}{4 \sqrt{2}}+O\left(\beta\mathcal{J}\right)^3\right)\frac{1}{\sqrt{q}}~,\\
b_2/\mathcal{J} &=\left(\sqrt{2}-\frac{\left(\beta\mathcal{J}\right) ^2}{4 \sqrt{2}}+O\left(\beta\mathcal{J}\right)^3\right)+ \left(-\frac{1}{\sqrt{2}}+\frac{5\left(\beta\mathcal{J}\right) ^2}{8 \sqrt{2}}+O\left(\beta\mathcal{J}\right)^3\right)\frac{1}{q}~,\\
b_3/\mathcal{J} &=\left(\sqrt{6}-\frac{\sqrt{3}\left(\beta\mathcal{J}\right) ^2}{4\sqrt{2}}+O\left(\beta\mathcal{J}\right)^3\right)+ \left(-\frac{5}{\sqrt{6}}-\frac{13\left(\beta\mathcal{J}\right) ^2}{8 \sqrt{6}}+O\left(\beta\mathcal{J}\right)^3\right)\frac{1}{q}~,\\
b_4/\mathcal{J} &=\left(2 \sqrt{3}-\frac{\sqrt{3}\left(\beta\mathcal{J}\right) ^2}{4}+O\left(\beta\mathcal{J}\right)^3\right)+\left(-\frac{35}{6 \sqrt{3}}+\frac{85\left(\beta\mathcal{J}\right) ^2}{48 \sqrt{3}}+O\left(\beta\mathcal{J}\right)^3\right)\frac{1}{q} ~,\\
b_5/\mathcal{J} &= \left(2 \sqrt{5}-\frac{\sqrt{5}\left(\beta\mathcal{J}\right) ^2}{4}+O\left(\beta\mathcal{J}\right)^3\right)+ \left(-\frac{377}{36 \sqrt{5}}-\frac{883\left(\beta\mathcal{J}\right) ^2}{288 \sqrt{5}}+O\left(\beta\mathcal{J}\right)^3\right)\frac{1}{q}~.
\end{cases}
\end{equation}
Note that this is a double expansion, first in $1/q$, and then in $\beta \mathcal{J}$, that seems reliable for small $\beta \mathcal{J}$.

\begin{figure}[H]
        \centering
         \subfigure[]{
                \includegraphics[scale=0.5]{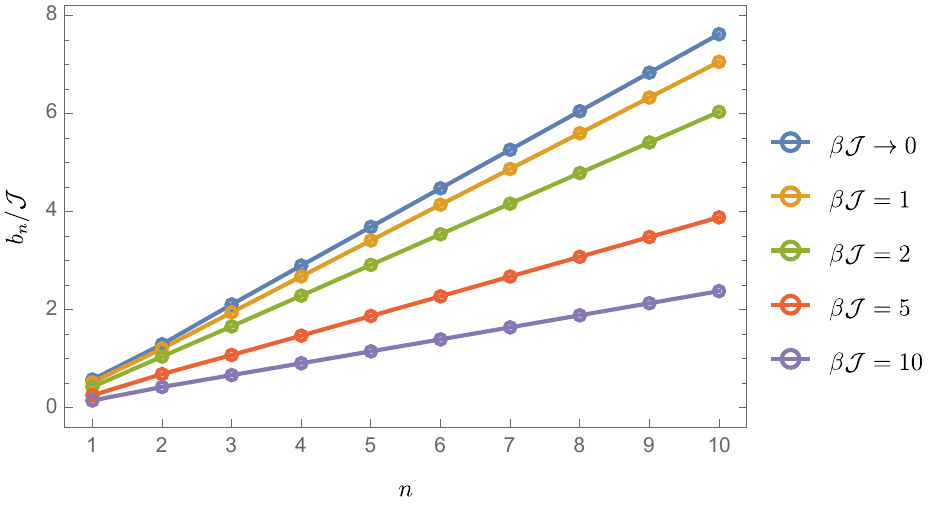}\label{finite_temp_large_q(a)}}  \quad\quad
       \subfigure[]{
                \includegraphics[scale=0.5]{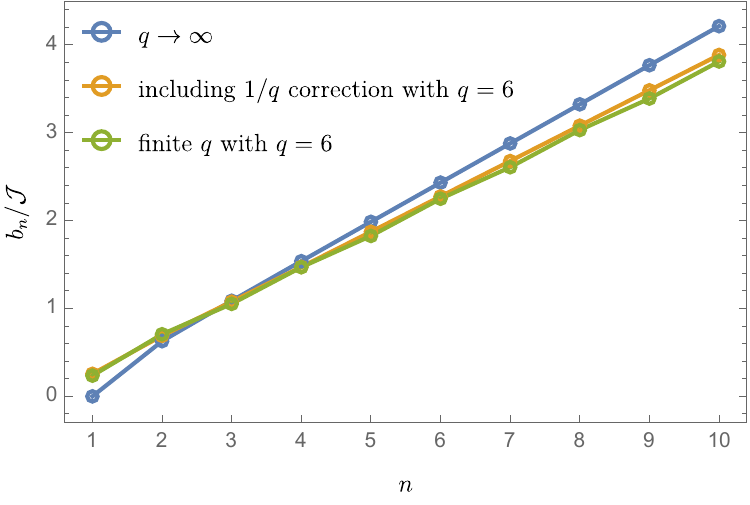} \label{large_q_finite_q_finite_b_comparison}}   
                
                \caption{Lanczos coefficients as a function of $n$, computed perturbatively in the large-$q$ expansion including the $1/q$ correction with $q=6$ at finite temperature. In (a) we plot different values of $\beta \mathcal{J}$ while in (b) we compare to the finite $q=6$ result at $\beta \mathcal{J}=5$.}\label{fig:finite_temp_large_q}
\end{figure}

We can also perform an expansion for large $\beta \mathcal{J}$, where we expect both the Lyapunov and the Krylov exponent to saturate the chaos bound. Perturbatively, for large $\beta \mathcal{J}$, they are given by,
\begin{equation}
\begin{cases}
b_1/\mathcal{J} &= \left(\frac{\sqrt{2}\pi }{\beta\mathcal{J}}- \frac{2\sqrt{2}\pi }{(\beta\mathcal{J})^2} +O\left(\frac{1}{\beta\mathcal{J}}\right)^{3}\right)\frac{1}{\sqrt{q}}~,\\
b_2/\mathcal{J} &=\left(\frac{\sqrt{2}\pi }{\beta\mathcal{J}} - \frac{2\sqrt{2}\pi }{(\beta\mathcal{J})^2} +O\left(\frac{1}{\beta\mathcal{J}}\right)^{3}\right)+\left(\frac{\sqrt{2} \pi}{\beta\mathcal{J} }+ \frac{\sqrt{2} \pi  \left(7 \pi ^2-6\right)}{9 (\beta \mathcal{J})^2} + O\left(\frac{1}{\beta\mathcal{J}}\right)^3\right)\frac{1}{q}~,\\
b_3/\mathcal{J} &=\left(\frac{\sqrt{6}\pi }{\beta\mathcal{J}} -\frac{2 \sqrt{6} \pi }{(\beta \mathcal{J})^2} +O\left(\frac{1}{\beta\mathcal{J}}\right)^{3}\right)+\left(\frac{\sqrt{3}\pi}{\sqrt{2}\beta\mathcal{J}}+ \frac{\sqrt{\frac{2}{3}} \pi  \left(3+7 \pi ^2\right)}{3 (\beta \mathcal{J})^2}+O\left(\frac{1}{\beta\mathcal{J}}\right)^3\right)\frac{1}{q}~,\\
b_4/\mathcal{J} &=\left(\frac{\sqrt{12}\pi }{\beta\mathcal{J}}-\frac{2 \sqrt{12} \pi }{(\beta \mathcal{J})^2}+O\left(\frac{1}{\beta\mathcal{J}}\right)^{3}\right)+\left(\frac{2\pi}{\sqrt{3}\beta\mathcal{J}}+ \frac{2 \pi  \left(6+7 \pi ^2\right)}{3 \sqrt{3} (\beta \mathcal{J})^2} +O\left(\frac{1}{\beta\mathcal{J}}\right)^{3}\right)\frac{1}{q}~,\\
b_5/\mathcal{J} &=\left(\frac{\sqrt{20}\pi }{\beta\mathcal{J}}-\frac{2 \sqrt{20} \pi }{(\beta \mathcal{J})^2}+O\left(\frac{1}{\beta\mathcal{J}}\right)^{3}\right)+\left(\frac{\sqrt{5}\pi }{2\beta\mathcal{J}}+ \frac{\sqrt{5} \pi  \left(15+14 \pi ^2\right)}{9 (\beta \mathcal{J})^2} + O\left(\frac{1}{\beta\mathcal{J}}\right)^{3}\right)\frac{1}{q}~.
\end{cases} \label{large_beta_large_q}
\end{equation}
The first 10 Lanczos coefficients up to order $1/q$ are shown for different temperatures in Fig.~\ref{fig:finite_temp_large_q}, where we also compare to numerical finite temperature results at finite $q$ (see section \ref{sec: Single SYK and saturation of Lyapunov exponent}), again finding good agreement.

\noindent \textbf{A comparison to the Lyapunov exponent including $1/q$ corrections.} We will now use the Lanczos coefficients found to compare the Krylov exponent with the Lyapunov exponent. The latter is only known perturbatively at large inverse temperatures. For any given $q$, it is possible to use the results in \cite{Maldacena:2016hyu} to obtain,

\begin{equation}\label{lambdaLprelim}
    \lambda_L = \frac{2\pi}{\beta} \left( 1 - \frac{48 \pi q^3 \cos ^2\left(\frac{\pi }{q}\right)}{2 \pi  (q-2) (q-1) \cos \left(\frac{2 \pi }{q}\right)-q ((q-6) q+6) \sin \left(\frac{2 \pi }{q}\right)}\frac{\alpha_S}{\beta \mathcal{J}} +\cdots \right) \,,
\end{equation}
where $\alpha_S$ is a $q$ dependent constant associated with the finite temperature corrections to the free energy, which can be fixed numerically. At large $q$, $\alpha_S$ has been computed analytically, including its first $1/q$ correction \cite{Tarnopolsky:2018env}, 
\begin{equation}
    \alpha_S = \frac{1}{4 q^2}-\frac{\pi ^2+12}{24 q^3} + \cdots \,.
\end{equation}
It is straightforward then to compute the correction to the Lyapunov exponent at large inverse temperatures in the large $q$ expansion, obtaining,\footnote{This is a good approximation for large $q$, but at finite $q$, it is not particularly precise. For instance, for $q=4$ and $q=6$, the exact correction to maximal chaos, see \eqref{lambda_C}, is given by $-4.28/\beta\mathcal{J}$ and $-3.11/\beta \mathcal{J}$, respectively, while this approximation gives $-3.04/\beta\mathcal{J}$ and $-2.69/\beta \mathcal{J}$. Alternatively, we can use the Pad\'e approximation for $\alpha_S$ in \eqref{lambdaLprelim} to obtain a better match for smaller values of $q$ \cite{Tarnopolsky:2018env}.}
\begin{equation}
     \lambda_L = \frac{2\pi}{\beta} \left( 1 - \left(2 + \frac{5 \pi ^2-12}{9 q} \right) \frac{1}{\beta \mathcal{J}} + \cdots \right) \,.
     \label{lyapunov_large_q_corr}
\end{equation}
The $2/\beta \mathcal{J}$ correction already comes from the leading term in the large $q$ expansion, where it was shown that both the Lyapunov and the Krylov exponent match at any temperature. The aim here is to see whether this match continues to hold to next order in $1/q$.

For this, we use the expansion at large inverse temperatures in \eqref{large_beta_large_q}. 
Note that here the Lanczos coefficients were obtained using the moment method. It is also possible to use this method without the explicit computation of the $\tau$-derivatives by using equations of motion.\footnote{The equation of motion for $g(\tau)$ is given in \eqref{diff_eqn}, while the one for $h(\tau)$ is given by \cite{Tarnopolsky:2018env}
\begin{equation}
    \partial^2_\tau h = 2 \mathcal{J}^2 e^g h +\frac{1}{2} \partial^3_\tau (g*g) - 2\mathcal{J}^2 e^g (g +\frac{1}{2} g^2) \,.
\end{equation}
Combining both equations of motion allows to compute the higher order moments $\partial^{2n}_\tau (g(\tau)/q + h(\tau)/q^2|_{\tau=\beta/2}$, with only having as an input the value of $g(\tau)$, $h(\tau)$ (and its first derivatives) at $\tau=\beta/2$.} We verified this gives the same results for all the $b_n$ we could compare. It is possible to guess a closed formula for the $b_n$, 
\begin{equation}
\begin{split}
    b_n  =  &\left( \frac{\pi  \sqrt{n (n-1)}}{\beta \mathcal{J}} - \frac{2 \pi  \sqrt{n (n-1)}}{(\beta \mathcal{J})^2} +\cdots \right)
    \\ &
 +\frac{1}{q} \left( \frac{\pi  \sqrt{\frac{n}{n-1}}}{\beta \mathcal{J}} + \frac{\frac{2\pi }{3} (2 n-5) \sqrt{\frac{n}{n-1}}+\frac{7}{9} \pi ^3 \sqrt{(n-1) n}}{(\beta \mathcal{J})^2} + \cdots \right) + \cdots \,,
 \end{split}
\end{equation}
that reproduces the $b_n$ in \eqref{large_beta_large_q} (for $n>1$) and though not shown for brevity, we also checked that it matches at least up to $n=8$. It is easy to extract the asymptotic form to compute the Krylov exponent, for which we obatin,
\begin{eqnarray} \label{krylov_large_q_corr}
    \lambda_K = \frac{2\pi}{\beta} \left( 1-\left( 2 - \frac{7 \pi ^2+12}{9 q} \right) \frac{1}{\beta \mathcal{J} } + \cdots \right) \,.
\end{eqnarray}
Note that there is a slight mismatch in the large $q$ correction between \eqref{lyapunov_large_q_corr} and \eqref{krylov_large_q_corr}. We find that the bound still holds as $\lambda_L \leq \lambda_K$, but it is not tight anymore,
\begin{equation}
   \frac{\beta}{2\pi} (\lambda_K - \lambda_L) = \frac{4 \pi ^2}{3 q \beta \mathcal{J} } + \cdots \,.
\end{equation}
In this case the mismatch is doubly suppressed, both in large $q$ and large inverse temperatures. In the next section we will study the single SYK at finite $q$ numerically, where, even if an analogous small difference might exist, we might not be able to detect it due to numerical errors. But we will study models in section \ref{sec:def_SYK} where deviations of order one exist between the Krylov and the Lyapunov exponent both at finite and infinite $q$.

\subsection{Finite $q$}\label{sec:single_finite_q}
\label{sec: Single SYK and saturation of Lyapunov exponent}

The final analysis we perform for a single SYK model is for finite $q\geq 4$ and finite temperatures. The aim is to find whether the conjectured bound \eqref{bound} holds and how tight it is. For $q\geq 4$ the autocorrelation function can only be obtained numerically. In order to get to the regime of very low temperatures where the maximal chaos bound is almost saturated, we need high numerical precision in our calculations. We find that it is necessary to get to at least $\beta \mathcal{J} \sim 100$, to be close to maximal chaos.

To find the Lanczos coefficients, one could proceed by solving the Schwinger-Dyson equations \eqref{field equations large N i} numerically and then taking numerical derivatives. This method however quickly accumulates errors. The reason is that at very low temperatures, $G(\tau)$ becomes very flat near $\tau \sim \beta/2$, which is the point at which derivatives need to be evaluated to compute the moments \eqref{momentdef} at finite temperature. 
Then, in order to get higher-$n$ Lanczos coefficients it is necessary to take further and further derivatives, whose values become smaller and smaller, and even small numerical errors become important.

Instead, it is more convenient to work in frequency space and numerically compute the spectral function $\rho(\omega)$ defined by\footnote{The expression $\lim_{\varepsilon\rightarrow 0}G(it+\varepsilon)$ should be interpreted as analytically continuing the positive $\tau>0$ branch of the Euclidean correlator as described in \cite{Maldacena:2016hyu}. This is as opposed to naively plugging $it+\varepsilon$ in the Euclidean correlator and performing the limit. This avoids the need to  interpret  $\text{sgn}(\tau)$ and $|\tau|$ for complex values of the parameter.}
\begin{equation}\label{eq:spectral_fct_main}
 \rho(\omega) = \frac{G^{>}(\omega)}{2\pi}(1+e^{-\beta\omega})~,\quad G^{>}(t) \equiv \frac{1}{N}\sum_{i}\langle \psi_i(t) \psi_i(0)\rangle_{\beta}=\lim_{\varepsilon\rightarrow 0}G(it+\varepsilon)~,
\end{equation}
using the procedure introduced in \cite{Sa:2021tdr}. See appendix \ref{app:spectral_fct} for details. From $\rho(\omega)$ we can find $G(\tau+\beta/2)$ by the following relation \cite{Maldacena:2016hyu}
\begin{equation}
    G(\tau+\beta/2) = \int d \omega\; e^{-\omega\left(\tau+\frac{\beta}{2}\right)} \frac{\rho(\omega)}{1+e^{-\beta \omega}} ~.\label{GWtau}
\end{equation}
Taking derivatives, we find that
\begin{equation}
\label{mus spectral}
\mu_{2n} = \frac{d^{2n}}{d{\tau}^{2n}}\left(2G(\tau+\beta/2)\right)|_{\tau=0} = \int d \omega\; \omega^{2n}e^{-\frac{\omega\beta}{2}} \frac{2\rho(\omega)}{1+e^{-\beta \omega}}~.
\end{equation}
We compute the moments $\mu_{2n}$ by numerically evaluating the integral in \eqref{mus spectral} and then obtain the Lanczos coefficients using the algorithm \eqref{coeffs from moments}. With this method we are able to reliably compute up to $\sim40$ Lanczos coefficients, with inverse temperatures as large as $\beta \mathcal{J} \sim 100$. Using this procedure, all these numerical computations can be done in the order of minutes on a standard laptop.

\

\begin{figure}[H]
\centering
                \includegraphics[scale=0.6]{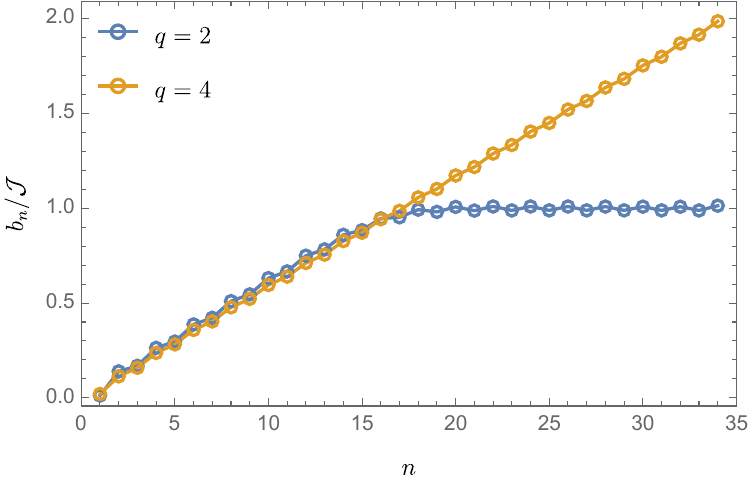}       
                \caption{First 35 Lanczos coefficients for the single $q=4$ (yellow) and $q=2$ (blue) SYK model at $\beta\mathcal{J}=50$. While the coefficients saturate for the $q=2$ model, they keep growing linearly for the $q=4$ one.}\label{fig:q_4_q_2_comparison2}
\end{figure}

The first thing we note is that the Lanczos coefficients for $q=4$, unlike those of $q=2$, do not saturate for large values of $n$ and finite inverse temperature. The comparison between the results in section \ref{sec:The_integrable_q2_model} and the Lanczos coefficients for the $q=4$ model at finite temperature can be found in Fig.~\ref{fig:q_4_q_2_comparison2}.

Note also that the even and the odd coefficients are staggered, but both sets grow linearly for large values of $n$ with equal slope and the staggering is a very small effect. To find $\lambda_{K}$, we calculate the slope from the largest 3 odd coefficients computed. We perform the same analysis for different values of $\beta \mathcal{J}$ to get the Krylov exponent $\lambda_K$ as a function of $\beta\mathcal{J}$.

\begin{figure}[H]	
    \centering
    \vspace{0mm}
    \includegraphics[width=0.5\columnwidth]{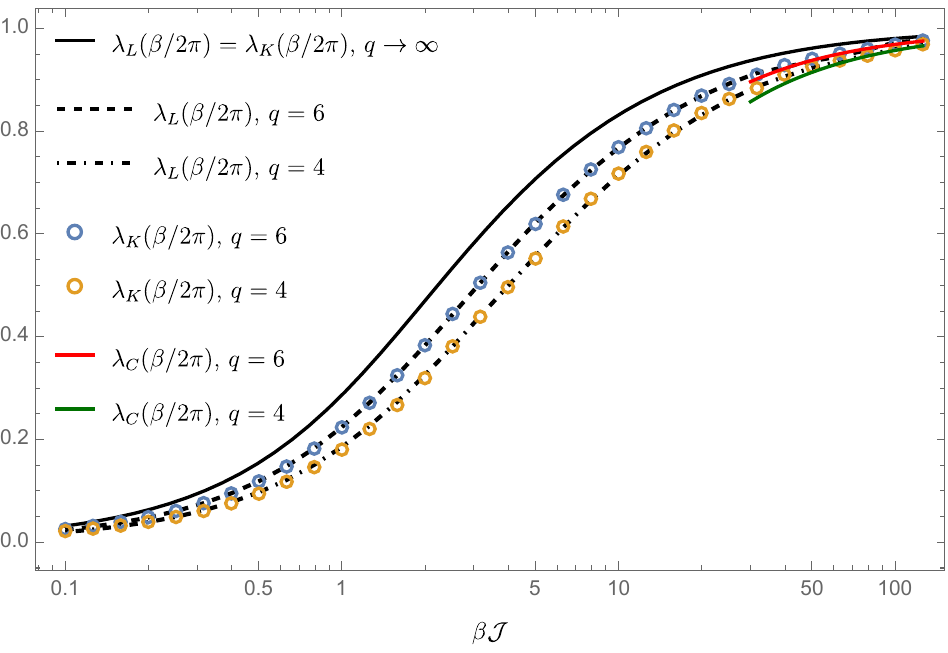} \caption{Krylov and Lyapunov exponent as a function of $\beta \mathcal{J}$ in the single SYK model. The dots are numerical computations of $\lambda_K$ for $q=4$ and $q=6$. The dashed-dotted and dashed curves are the respective Lyapunov exponents, computed numerically. The solid curves are analytical computations of the exponent in the large-$q$ limit (black), and analytic results at low temperatures for the Lyapunov exponent for $q=4$ (green) and $q=6$ (red). To the precision achieved, we find that $\lambda_K \simeq \lambda_L$ and the differences, while non-vanishing, fall within our numerical accuracy.}\label{fig:q4_q6_q_inf_lyapunov_krylov}
\end{figure}

In Fig.~\ref{fig:q4_q6_q_inf_lyapunov_krylov} we show the result of this computation both for $q=4$ and $q=6$. We also plot their corresponding Lyapunov exponents, that we also computed numerically following the procedure described in appendix \ref{sec: comp Lyapunov}. For comparison, we also include the analytic expression for the Lyapunov exponent computed in the conformal (low temperature) limit and its leading correction \eqref{lambda_C}. Finally, we also include the analytic curve for the Krylov exponent at large-$q$ which is known to be the same as the Lyapunov exponent and given by \eqref{large_q_krylov}. As can be seen in this plot, we find $\lambda_{K} \simeq \lambda_{L}$ for both $q=4$ and $q=6$. It is worth noting that we do not necessarily expect the two exponents to coincide precisely, given our results of section  \ref{sec:single_large_q_corr} for the large $q$ expansion which already showed a small disagreement between the two.  However, the differences we find between the Krylov and Lyapunov exponents at finite $q$ fall
within our numerical accuracy, preventing us from drawing further conclusions about them. Nonetheless, while any differences would be very small here, we will see in the next section that in deformed SYK models the difference between the two exponents can become rather large.

\section{Krylov complexity of deformed SYK models} \label{sec:def_SYK}

In this section, we will study deformed SYK models in which the total Hamiltonian is the sum of two SYK Hamiltonians with different number of fermions $q$ and $\tilde{q}$ in each interaction term, see section \ref{subsec:def_SYK}. At finite temperature, these Hamiltonians generate interesting renormalisation group (RG) flows \cite{Anninos:2022qgy}. As a consequence, the Lyapunov exponent does not behave monotonically as a function of $\beta \mathcal{J}$. We are interested in comparing the behaviour of the Krylov exponent $\lambda_K$ along the RG flow to that of the Lyapunov exponent $\lambda_L$.

When $\tilde{q} \neq 2$, the thermal RG flow interpolates two regions of nearly maximal chaos. Between these regions, the Lyapunov exponent decreases and then increases again when approaching the second maximally chaotic region. We refer to these models as chaos-to-chaos RG flows and we study them in section \ref{sec:CC_RG}, both at infinite and finite $q, \tilde{q}$.

When $\tilde{q}=2$, the intermediate near fixed point is still near-maximally chaotic, but the deformation consists of an integrable Hamiltonian. The Lyapunov exponent decreases at sufficiently large $\beta \mathcal{J}$. We refer to this case as chaos-to-integrable RG flow, and we study it both at finite and infinite $q$ in section \ref{sec:CI_RG}.

\subsection{Chaos-to-chaos RG flows}\label{sec:CC_RG}
\subsubsection{Large $q,\tilde{q}$}
\label{sec:Chaos-to-chaos flows}

The simplest model that presents non-monotonic behaviour of the Lyapunov exponent is the deformed SYK model with Hamiltonian \eqref{deformed Hamiltonian} and $\tilde{q}=q/2$, in the large-$q$ limit. The two-point correlator of this model is known analytically for any value of $\beta \mathcal{J}$ and $s$, see \eqref{largeq1}, \eqref{largeqcon} \cite{Jiang:2019pam, Anninos:2020cwo, Anninos:2022qgy}.
If $s \ll 1$, the infrared of the theory is known to have two regions where the Lyapunov exponent is nearly maximal, see \eqref{lyapunov_def}. In between these two regions, the Lyapunov exponent is sub-maximal, but it can nevertheless be computed numerically \cite{Jiang:2019pam}, see details in appendix \ref{sec: comp Lyapunov}. Since we have an analytic expression for the two-point function, we can compute $\lambda_K$ using both methods described in sections \ref{sec: lanczos from moments} and \ref{sec: slope from pole}. It is also possible to use the pole method to compute $\lambda_K$ numerically in this model \cite{cao-nandy}.

We start by computing the asymptotic slope of the Lanczos coefficients by looking at the pole of the Wightman autocorrelation function. It is straightforward to verify that the pole $\tau_*$ of $G(\tau + \beta/2)$ in \eqref{largeq1} is given by, 
\begin{equation}
    \tau_*=\frac{\beta  \cos ^{-1}\left(-\frac{s^2(\beta\mathcal{J})^2}{\sqrt{(\beta\mathcal{J})^2\nu^2+s^4(\beta\mathcal{J})^4}}\right)}{2  \nu },
\end{equation}
where $\nu$ is implicitly given as a function of $\beta \mathcal{J}$ and $s$ in \eqref{largeqcon}. This immediately gives,
\begin{eqnarray}
    \lambda_K = \frac{2 \pi \nu}{\beta  \cos ^{-1}\left(-\frac{s^2(\beta\mathcal{J})^2}{\sqrt{(\beta\mathcal{J})^2\nu^2+s^4(\beta\mathcal{J})^4}}\right)},
\end{eqnarray}
that reproduces the single SYK large-$q$ result \eqref{large_q_krylov} when $s\to 0$. We emphasise again that this result is valid for any value of $\beta \mathcal{J}$ and $s$. If $s \ll 1$, we can further expand $\nu$ analytically at low temperatures close to each of the near fixed points, see \eqref{nu_int} and \eqref{nu_deep}. This gives,
\begin{equation} \label{krylov_exp_def}
    \lambda_K = \begin{cases} \frac{2\pi}{\beta} \left( 1 -\frac{2}{\beta  \mathcal{J}}+
 \frac{4}{(\beta \mathcal{J})^2}-\frac{24+\pi ^2}{3 (\beta \mathcal{J})^3} +\frac{2 (\beta \mathcal{J})^2 s^2}{(\beta \mathcal{J})^3}+ \cdots \right)
 \,, & \text{Intermediate IR} \,, \\
 \frac{2\pi}{\beta} \left( 1 - \frac{\sqrt{4 s^2+1}-1}{\beta  \mathcal{J} s^2}
 +\frac{2+4 s^2-2 \sqrt{4 s^2+1}}{(\beta \mathcal{J})^2 s^4}+ 
  \cdots
 \right)
 \,, & \text{Deep IR} \,,
\end{cases}
\end{equation}
where we observe \emph{for the first time} an SYK-like model where the Krylov exponent $\lambda_K$ does not provide a tight bound for the Lyapunov exponent $\lambda_L$. Comparing to \eqref{lyapunov_def}, we see that nevertheless, the conjectured bound \eqref{bound} still holds 
around the fixed points, where the expansion is valid. Note that the first four terms in the intermediate IR expansion, are exactly the ones of a single SYK, see \eqref{single_SYK_krylov_expansion}, so the first correction that depends on $s$ will be very suppressed. Furthermore, the corrections away from maximal chaos in the deep IR are extremely small for small $s$.

\begin{figure}[H]
        \centering
         \subfigure[$s=1$]{
                \includegraphics[scale=0.45]{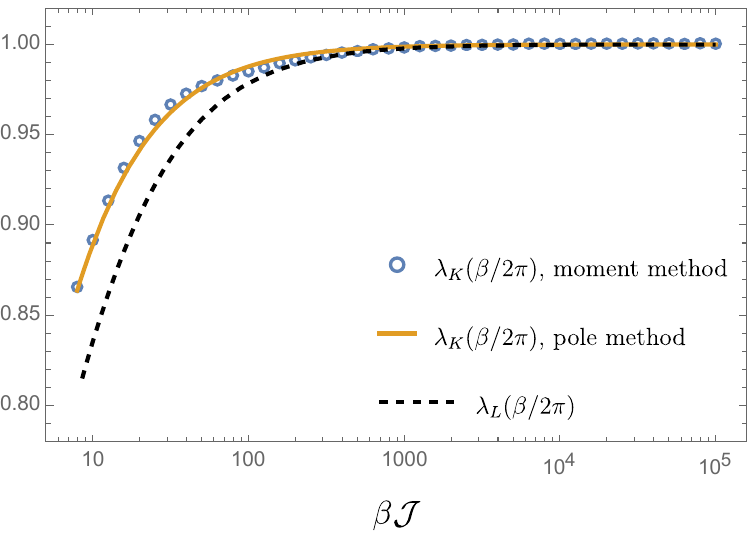}\label{s11}}  \quad\quad
       \subfigure[$s=0.1$]{
                \includegraphics[scale=0.45]{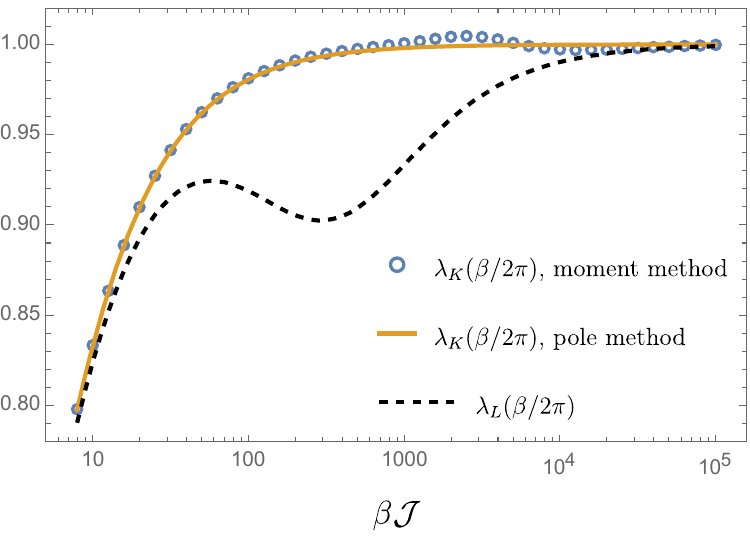} \label{s0p11}}   
       \subfigure[$s=0.01$]{
                \includegraphics[scale=0.45]{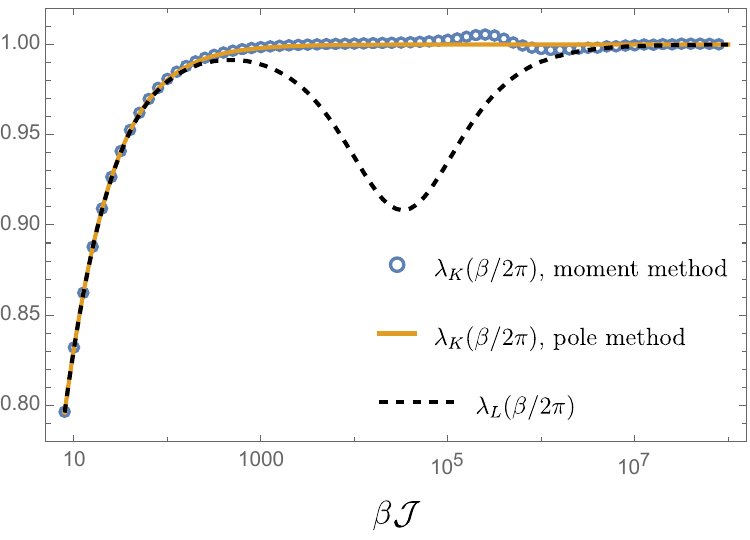}\label{s0p01}}  \quad\quad
       \subfigure[$s=0.001$]{
                \includegraphics[scale=0.45]{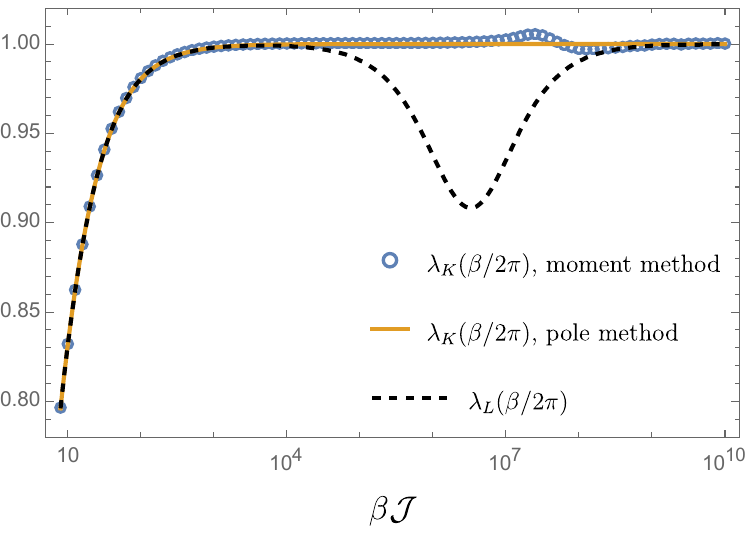} \label{s0p001}}                           
                \caption{Krylov and Lyapunov exponents as a function of $\beta\mathcal{J}$ for the deformed model \eqref{deformed Hamiltonian} in the large-$q$ limit with $q=2\tilde{q}$. The Krylov exponent is computed in two ways: the blue dots are computed using the moments method described in section \ref{sec: lanczos from moments} whilst the solid orange lines are computed from the pole of the autocorrelation function. The black dashed lines show the Lyapunov exponent, which is computed numerically. Plots are shown for different values of $s$. Unlike the Lyapunov exponent, the Krylov exponent grows monotonically and does not detect the region of sub-maximal chaos between the near maximally chaotic regions.}\label{fig:lyapunov}
\end{figure}

In order to have an independent calculation of the Krylov exponent, we also compute the slope of the Lanczos coefficients using the moment method \eqref{coeffs from moments}. We indeed verify that the slope is linear and the Lanczos coefficients do not stagger for the values of $n$ that are computationally available, which validates the use of the pole method above. 
The full plot of the Krylov exponent as a function of $\beta \mathcal{J}$ for different values of $s$ can be found in Fig.~\ref{fig:lyapunov}. We include the computation of $\lambda_K$ using both methods, as well as the result of the Lyapunov exponent, computed numerically, as described in appendix \ref{sec: comp Lyapunov}.

The first aspect to notice is that both methods agree to good precision, except for a small bump that can be seen in the computation using the moments method for inverse temperatures slightly larger than $\beta \mathcal{J} \sim s^{-2}$. If physical, this bump would not violate the left inequality in \eqref{bound}, but it would violate the right one. Moreover, it would mean that the two methods of computing the Krylov exponent are not equivalent. However, we checked that this bump is not physical. In particular, it is an artifact of the numerical procedure to compute the moments at large Lanczos coefficient index $n$, and we verified that the size of the bump decreases with $n$. See appendix \ref{app:bump}.\footnote{We are grateful to Xiangyu Cao and Pratik Nandy for discussions on this point.}

\begin{figure}[H]
        \centering
         \subfigure[$s=1$]{
                \includegraphics[scale=0.45]{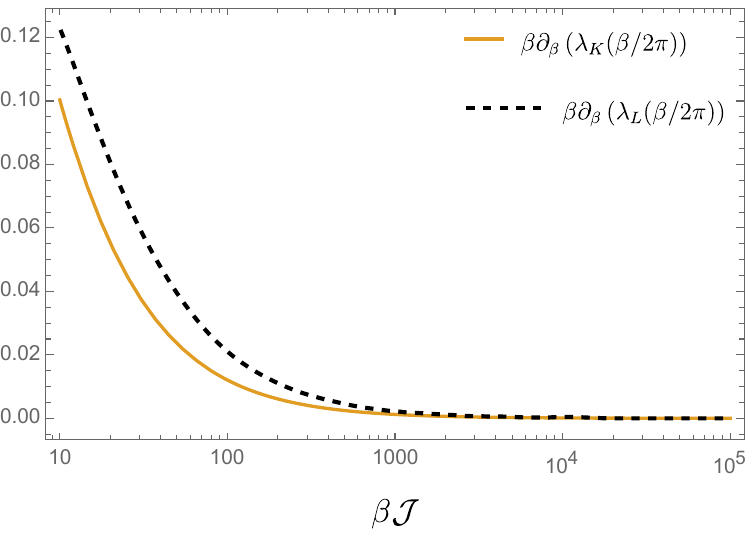}\label{partial_s1}}  \quad\quad
       \subfigure[$s=0.1$]{
                \includegraphics[scale=0.45]{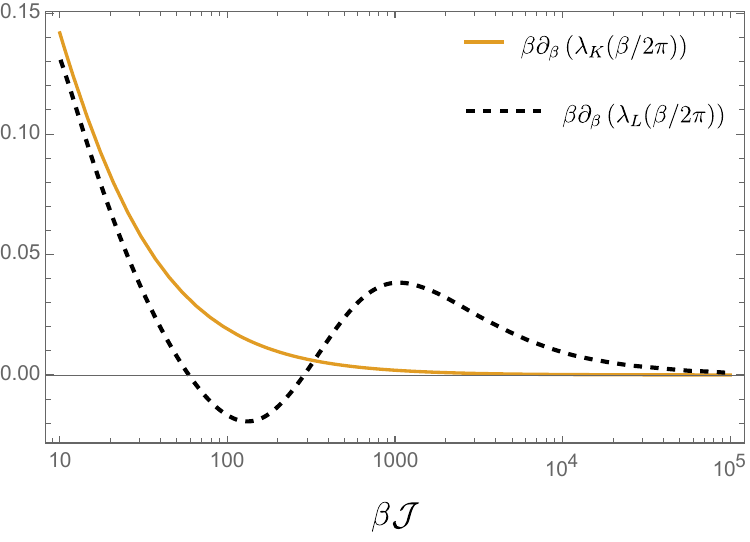} \label{partial_s0p1}}   
       \subfigure[$s=0.01$]{
                \includegraphics[scale=0.45]{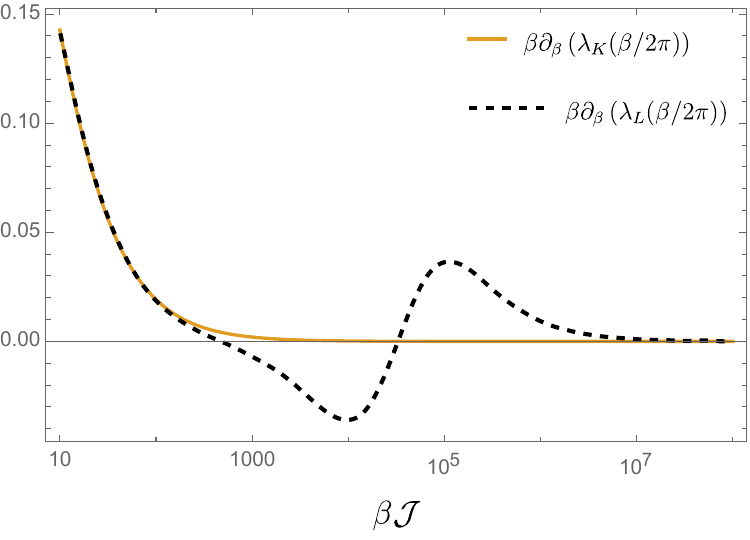}\label{partial_s0p01}}  \quad\quad
       \subfigure[$s=0.001$]{
                \includegraphics[scale=0.45]{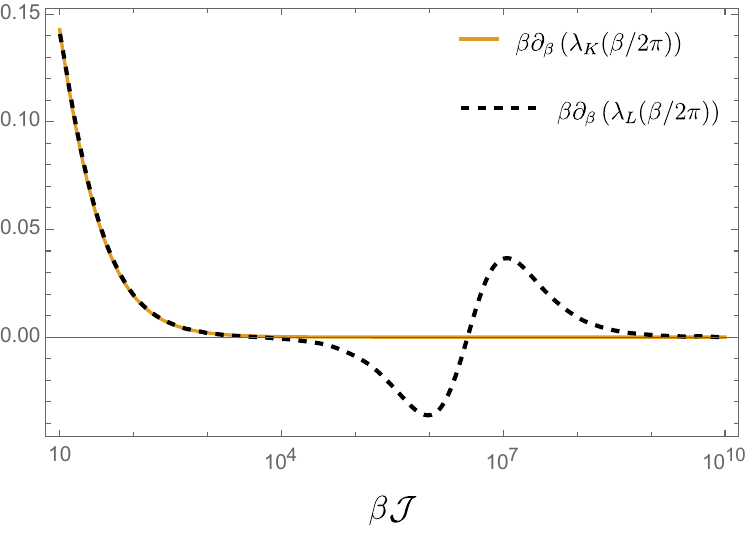} \label{partial_s0p001}}                           
                \caption{Derivatives of Krylov and Lyapunov exponents as a function of $\beta\mathcal{J}$ for the deformed SYK in the large-$q$ limit with $q=2\tilde{q}$. We observe that in all cases $\beta\partial_\beta \left(\lambda_{K}(\beta/2\pi)\right) \geq 0$, whilst this is not the case for the Lyapunov exponent.}\label{fig:partial_lyapunov}
\end{figure}

Secondly, we observe that the Krylov exponent is always larger than or equal to the Lyapunov exponent, so that the bound between the two still holds for all the values of $\beta \mathcal{J}$ and $s$ explored. The difference with the single SYK is that now the two exponents differ significantly from each other. In fact, while the Lyapunov exponent exhibits non-monotonic behaviour and sub-maximal chaos in between the near-maximally chaotic regions, the Krylov exponent is monotonic in $\beta \mathcal{J}$. This can be clearly seen from Fig.~\ref{fig:partial_lyapunov}, where we plot $\beta\partial_\beta \left(\lambda_{K,L}(\beta/2\pi)\right)$ and observe that $\beta\partial_\beta \left(\lambda_{K}(\beta/2\pi)\right) \geq 0$. 

\textit{En passant}, we note that if instead of looking at the asymptotic growth of the Lanczos coefficients at large $n$, we focus on the approximately linear slope of the first few coefficients, we surprisingly find a non-monotonic behaviour that resembles closely the one of the Lyapunov exponent. We refer the reader to appendix \ref{app:enpassant} for plots substantiating the relation. It would be interesting to understand the physical reason for this feature, if any, in future work. It would also be interesting to understand how the various features presented in this section depend on $\mathfrak{n}=q/\tilde q$ and we leave this for future work.

\subsubsection{Finite $q,\tilde{q}$}

\label{sec:deformed_finite_q_ch_ch}
Next, we move to the study of deformed SYK models at finite $q$ and $
\tilde{q}$. The numerical methods used in section \ref{sec:single_finite_q} to compute the Krylov and the Lyapunov exponent, can be readily adapted to the deformed Hamiltonians \eqref{deformed Hamiltonian}.

To compute the Krylov exponent, we will be using the moments method, as in the single SYK at finite $q$. In contrast to the analytic control of the large-$q$ models, in the current case, numerical errors pose a challenge to a reliable computation of the autocorrelation function (and its moments) at very low temperatures. Using the techniques described in appendices \ref{sec: comp Lyapunov} and \ref{app:spectral_fct}, we managed to reliably compute the Krylov and Lyapunov exponents along the RG flow for inverse temperatures as high as $\beta \mathcal{J} \sim 100$. As we will see, this will not allow us to observe the full RG flow in the deep infrared, but will be enough to observe the main chaotic features of the flow for values of $s \sim 0.1$.

\

\begin{figure}[H]	
    \centering
    \vspace{0mm}
    \includegraphics[width=0.4\columnwidth]{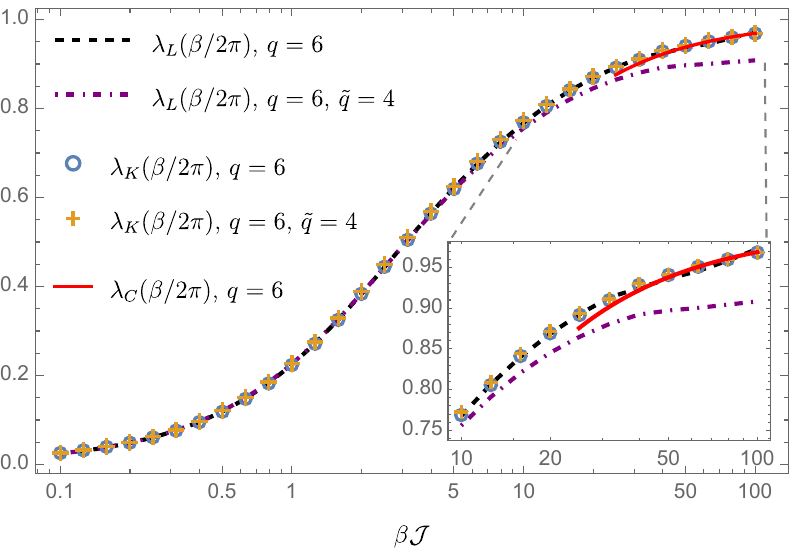} \caption{The Krylov exponent $\lambda_K$ and the Lyapunov exponent $\lambda_L$ as a function of $\beta \mathcal{J}$ for the deformed SYK model with $q=6$, $\tilde{q}=4$ and $s=0.2$. We added the analogous computations for the single SYK model with $q=6$ (together with the analytic expansion for the Lyapunov exponent at large $\beta \mathcal{J}$) as a reference (red). In the caption, we zoom into the region of large $\beta \mathcal{J}$, to clearly distinguish the behaviour of $\lambda_K$ from $\lambda_L$.}\label{fig:chaos_chaos}
\end{figure}

If we want to consider models that are maximally chaotic in the deep infrared, then $\tilde{q}$ has to be at least 4, which also means that $q\geq 6$. Reaching large values of $\beta\mathcal{J}$ is even more challenging in this case, since the spectral function $\rho(\omega)$ becomes sharply peaked at lower values of $\beta\mathcal{J}$ as we increase $q$, see appendix \ref{app:spectral_fct}. Nevertheless, we managed to compute both the Krylov and the Lyapunov exponents up to $\beta \mathcal{J}=100$ for $s=0.2$. The result can be seen in Fig.~\ref{fig:chaos_chaos}. From the results at large $q$, we expect that the Lyapunov exponent will initially grow for small $\beta \mathcal{J}$. At some intermediate region, this growth will stop, the Lyapunov exponent will decrease and eventually it will go back to near maximal chaos at very large $\beta \mathcal{J}$. At the computationally available inverse temperatures, we managed to clearly observe the slowdown of the growth of the Lyapunov exponent, see the dashed-dotted purple curve in Fig.~\ref{fig:chaos_chaos}. On the other hand, the Krylov exponent of the flow SYK follows closely the Krylov exponent of the single SYK with $q=6$ at all available inverse temperatures, until it saturates to the value of the near-conformal Lyapunov exponent result of a single $q=6$ SYK model \eqref{lambda_C}.

Once again, paralleling the large-$q$ results, the Krylov exponent shows a monotonic behaviour and is not able to follow the slowdown of the Lyapunov exponent. This does not violate either of the bounds in \eqref{bound}, but shows that the Krylov exponent does not provide a tight bound for chaotic-to-chaotic thermal RG flows in SYK. We therefore suspect that it cannot distinguish the different regimes of chaos-to-chaos RG flows.

\subsection{Chaos-to-integrable RG flows}\label{sec:CI_RG}

\subsubsection{Finite $q$ and  $\tilde{q}=2$}
\label{sec:deformed_finite_q_ch_int}

We now consider the deformed SYK model with $\tilde{q}=2$ and finite $q\geq4$. For $q=4$, the Lyapunov exponent has been numerically computed before showing an initial increase at low temperatures, with a later decrease at even lower temperatures \cite{Garcia-Garcia:2017bkg}. It is a subject of debate whether the Lyapunov exponent (in the large-$N$ limit) actually becomes zero at finite temperature or if this is only achieved in the strict zero temperature limit \cite{Garcia-Garcia:2017bkg, Kim:2020mho, Garcia-Garcia:2020dzm}.

\begin{figure}[H]
        \centering
         \subfigure[$q=6,\; \tilde{q}=2$]{
                \includegraphics[scale=0.5]{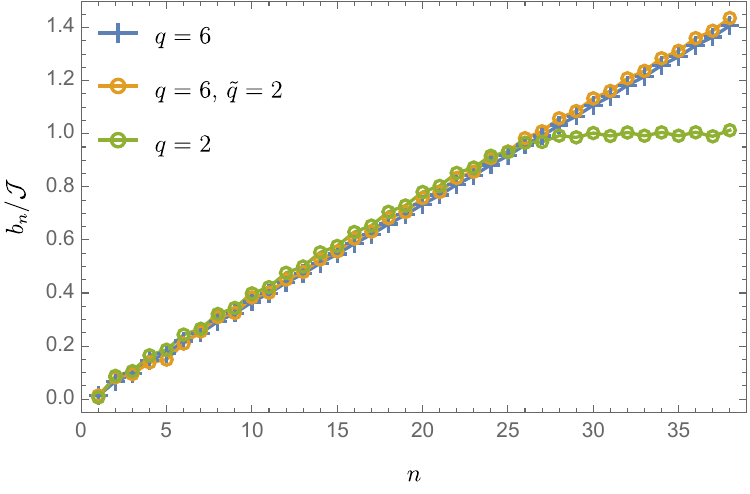}\label{b80_s1}}  \quad\quad
       \subfigure[$q=4,\; \tilde{q}=2$]{
                \includegraphics[scale=0.5]{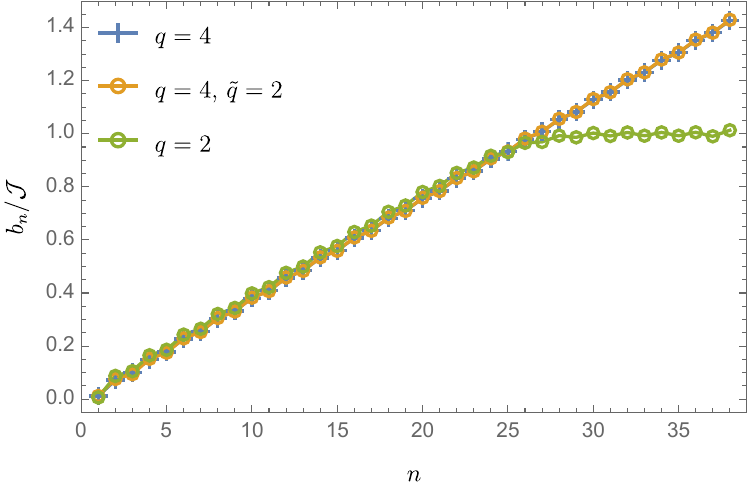} \label{b80_s0p1}}   
                
                \caption{First 38 Lanczos coefficients for the deformed SYK Hamiltonian with $s=0.2$ and $\beta\mathcal{J}=80$. For reference, we also include the first Lanczos coefficients for the single $q=2$ SYK (which saturate at large $n$) and those of the single $q$ model with $q=6$ in plot (a) and $q=4$ in plot (b).}\label{fig:fig:chaos_integrable_b_80}
\end{figure}

In order to compute the Krylov exponent, we first compute the Lanczos coefficients for fixed values of $\beta \mathcal{J}$ and $s$, using the numerical moments method. As an example, we show the first $\sim 40$ Lanczos coefficients for $\beta \mathcal{J}=80$ and $s=0.2$ in Fig.~\ref{fig:fig:chaos_integrable_b_80}. We compare them with the Lanczos coefficients of the single SYK with $q=2$, see section \ref{sec:The_integrable_q2_model}, and we clearly see that the saturation present in the $q=2$ model is not present in the deformed SYK model. Moreover, the Lanczos coefficients seem to closely follow those of the single SYK with $q$ fermions, both for $q=6$ and $q=4$, so it seems unlikely that the Krylov exponent will  decrease (when taking larger values of the Lanczos index $n$) for this choice of parameters. 

\begin{figure}[H]
        \centering
         \subfigure[$q=6,\; \tilde{q}=2$]{
                \includegraphics[scale=0.5]{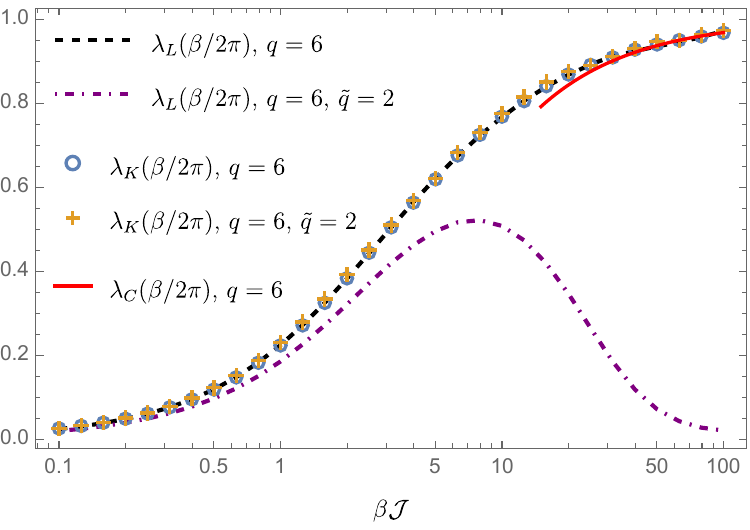}\label{s1}}  \quad\quad
       \subfigure[$q=4,\; \tilde{q}=2$]{
                \includegraphics[scale=0.5]{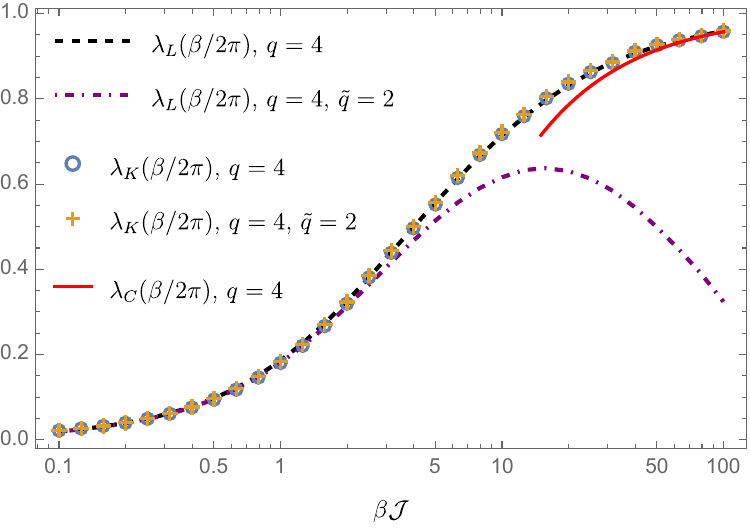} \label{s0p1}}   
                
                \caption{The Krylov exponent $\lambda_K$ and the Lyapunov exponent $\lambda_L$ as a function of $\beta \mathcal{J}$ for the deformed SYK model with $s=0.2$. We added the analogous computations for the single SYK model (together with its analytical prediction at large $\beta \mathcal{J}$) as a reference. At low temperatures $\lambda_L \leq \lambda_K$, but the bound is not tight. }\label{fig:chaos_integrable}
\end{figure}

\begin{figure}[H]
        \centering
         \subfigure[$q=6,\; \tilde{q}=2$]{
                \includegraphics[scale=0.5]{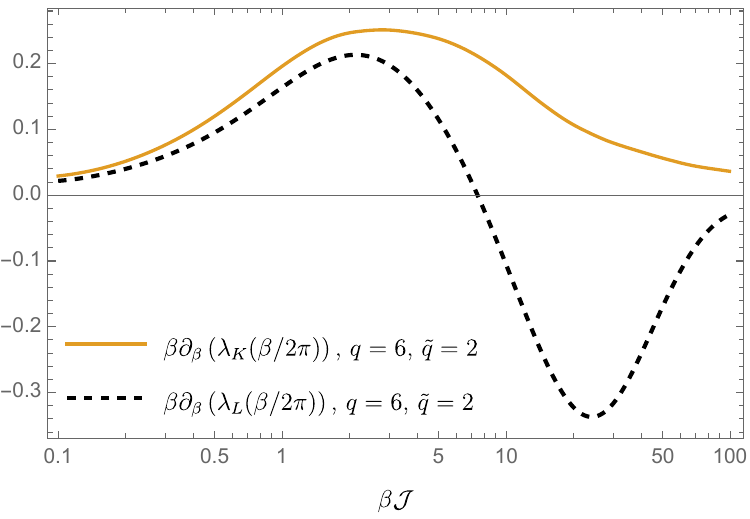}\label{s1_p}}  \quad\quad
       \subfigure[$q=4,\; \tilde{q}=2$]{
                \includegraphics[scale=0.5]{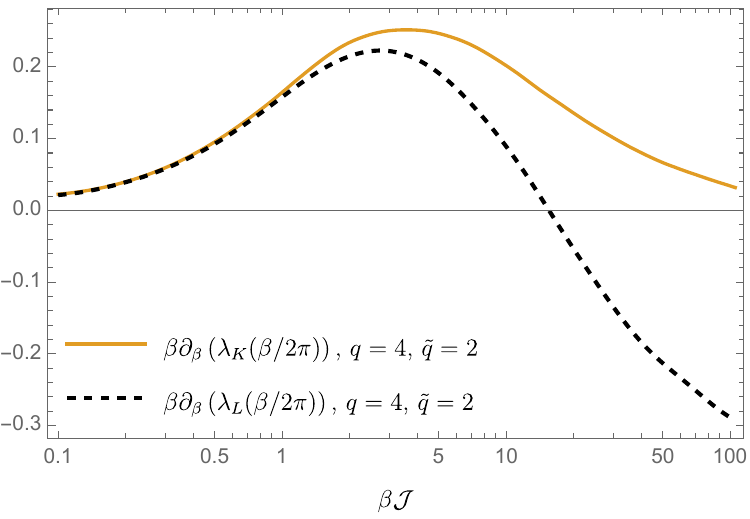} \label{s0p1_p}}   
                
                \caption{Derivatives of Krylov and Lyapunov exponents as a function of $\beta\mathcal{J}$ for the deformed model with $s=0.2$. Plots are shown for different values of $s$. We observe that $\beta\partial_\beta \left(\lambda_{K}(\beta/2\pi)\right) \geq 0$, whilst this is not the case for the Lyapunov exponent.}\label{fig:chaos_integrable_p}
\end{figure}

We show the full behaviour of the Krylov exponent as a function of $\beta \mathcal{J}$ for fixed $s=0.2$ in Fig.~\ref{fig:chaos_integrable}, both for $q=6$ and $q=4$ to $\tilde q=2$ flows, along with their corresponding Lyapunov exponents, computed using techniques in appendix \ref{sec: comp Lyapunov} . In the range of temperatures numerically available, we can clearly see a pronounced decrease of the Lyapunov exponent, which becomes even more evident in the $q=6$ case. In contrast, the Krylov exponent does not depart from the Krylov exponent of the single $q$ SYK model, that we also plot for reference. In particular, while the bound \eqref{bound} is still obeyed at all temperatures, we do not observe a decrease in the Krylov exponent, that monotonically increases towards the maximal chaos bound at very low temperatures. See also Fig.~\ref{fig:chaos_integrable_p}, where we plot $\beta \partial_\beta (\lambda_K \beta/2\pi)$. This is consistent with the behaviour found for $\lambda_K$ in the other examples considered so far, both in the single SYK and in the chaos-to-chaos RG flows.

Before going to a large-$q$ example of chaos-to-integrable flows, let us make a brief comment on the behaviour of the Lyapunov exponent at low temperatures. 
In \cite{Kim:2020mho}, it was reported that for the $q=4$ to $\tilde{q}=2$ SYK flow, the decay of the Lyapunov exponent at large inverse temperatures for large $s$ follows a power-law behaviour, $\lambda_L (\beta/2\pi) \sim (\beta \mathcal{J})^{-1}$. From \ref{s1}, it seems that the $q=6$ model with small $s$ also exhibits a power-law behaviour. 
In this regime, we observe that the tail at large $\beta \mathcal{J}$ seems to go as $\lambda_L (\beta/2\pi) \sim (\beta \mathcal{J})^{-2}$, which suggests that, as in \cite{Kim:2020mho}, at large $N$ the Lyapunov exponent $\lambda_L$ reaches zero only at infinite $\beta \mathcal{J}$. The precise low-temperature behaviour seems to depend both on $q$ and $s$. It would be interesting to understand this dependence better. 

\subsubsection{Large $q$ and $\tilde{q}=2$} \label{sec:largeq_qtilde2}
There is one more example of a chaos-to-integrable RG flow, that can be solved analytically, at least perturbatively in the deformation parameter. This is the model with $q \to \infty$ and $\tilde{q}=2$. We need to be careful when taking the large-$q$ limit of this model. The relevant Schwinger-Dyson equation from \eqref{SD2}, for $\tilde{q}=2$ becomes 
\begin{equation}
     \Sigma(\tau_1,\tau_2) =\mathcal{J}^2 \left(\frac{2^{q-1}}{q}  G(\tau_1,\tau_2)^{q-1}  +  s^{2} G(\tau_1,\tau_2)\right) ~.
\end{equation}
As can be seen, the deformation would become dominant if we take the large-$q$ limit naively. Instead, we re-scale the coupling $s^2 \to \s^2/q$ so that in the large-$q$ limit, using \eqref{large q 1st order}, we obtain,
\begin{eqnarray} \label{eq: large_q_qtilde_eom}
    \partial_{\tau}^{2}g(\tau) =  \mathcal{J}^2 \left(2 e^{g(\tau)} + \mathfrak{s}^2\right)~. 
\end{eqnarray}
This equation does not have an analytic solution, but can be solved both perturbatively in $\s$ \cite{Garcia-Garcia:2017bkg} or numerically. For the perturbative solution, we expand $g(\tau)$ as,
\begin{align} \label{large_q_integrable}
    g(\tau) = g^{(0)}(\tau)+\s^2 ~g^{(1)}(\tau)+O((\beta \mathcal{J} \s)^4) \,.
\end{align}
The undeformed solution is given by \eqref{1st correction large q}, while $g^{(1)}(\tau)$ is given by \cite{Garcia-Garcia:2017bkg}, 
\begin{align}
g^{(1)}(\tau)= \left(\frac{\beta \mathcal{J}}{2\nu}\right)^2
&\Big[\alpha\left(\frac{\tau}{\beta}\right)\tan{\left(\frac{\tau}{\beta}\right)}+\log{\cos{\left(\frac{\tau}{\beta}\right)}}+\frac{\tau}{\beta}\tan{\left(\frac{\tau}{\beta}\right)}+B(\nu)\left(\frac{\tau}{\beta}\tan{\left(\frac{\tau}{\beta}\right)}+1\right) \Big] \,,
\end{align}
with 
\begin{equation}
\begin{cases}
\alpha(x) &= \int^x\mathrm{d} t~\log\cos{t}=\frac{i}{2}  \text{Li}_2\left(-e^{2 i x}\right)+\frac{i x^2}{2}-x \log \left(1+e^{2 i
   x}\right)+x \log \cos (x) \,,\\
B(\nu) &=-\frac{-\alpha \left(-\nu\right) \tan
  \nu+\alpha \left(\nu
  \right) \tan \nu+2  \nu 
   \tan \nu+2 \log \cos
   \nu}{2  \nu  \tan
   \nu+2}\,.
   \end{cases}
\end{equation}
The perturbative correction to the Lyapunov exponent to this order was computed in \cite{Garcia-Garcia:2017bkg}, and it is given by
\begin{equation}
\label{lambda pert}
    \lambda^{\text{(pert.)}}_L = \frac{2\pi}{\beta} \left(\frac{2\nu}{\pi} - \frac{\s^2 \nu}{2\pi}\frac{ B(\nu) +\frac{19}{18} - \log 2}{\cos^2 \nu}  + \cdots \right)  \,.
\end{equation}

In principle, given that we have the analytic two-point function, we should be able to use either the moments method or look at the pole of the Wightman autocorrelator to compute  $\lambda_K$. However, in this case, there are two difficulties. Since the solution is only perturbative in $\s$ and the pole might depend on $\s$, the pole method cannot be applied to the expansion directly since it will not capture the correct location of the pole in the two-point function. We could still use the moments method, but then the solution can only be trusted up to $\beta \mathcal{J} \sim 1/\s$, where we do not expect to see a great deviation of the Krylov exponent from the Lyapunov exponent.

\begin{figure}[H]	
    \centering
    \vspace{0mm}
    \includegraphics[width=0.8\columnwidth]{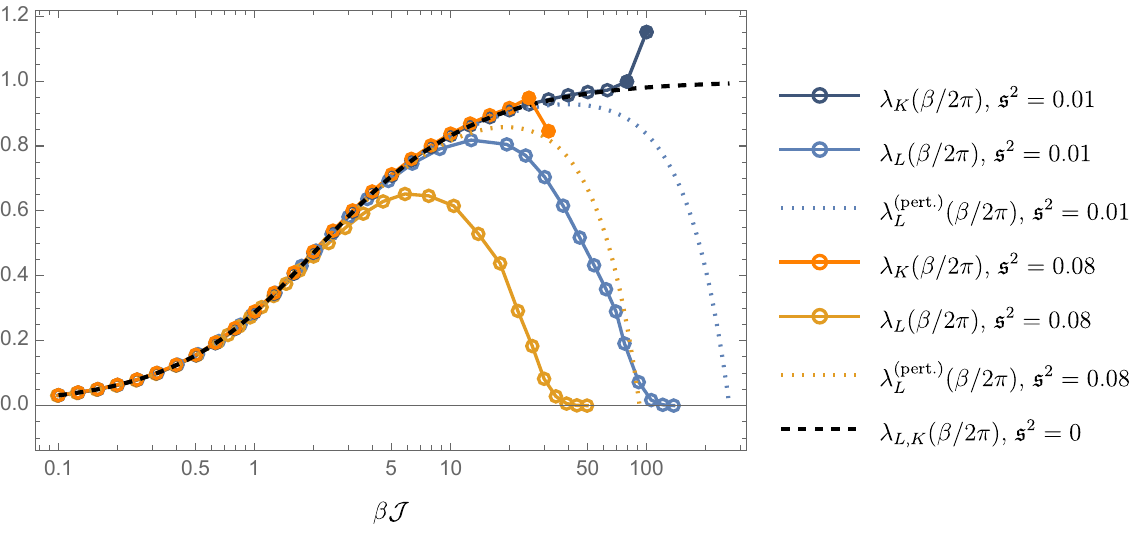} \caption{Krylov and Lyapunov exponents for the $q \to \infty$ and $\tilde{q}=2$ model. Plots are shown for $\s^2=0, 0.01, 0.08$. The solid curves are numerical computations. The blacked dashed curve is the analytical $\lambda_L = \lambda_K$ for $\s^2=0$, given by \eqref{large_q_krylov}. Note that for the Krylov exponent curves with $\s^2>0$, we indicate the first two points that are observed to deviate significantly from the $\s^2=0$ curve by filled circles. These points should not be taken as valid Krylov exponents, as the Lanczos coefficients are not linear. The dotted lines are the pertubative Lyapunov exponents computed from \eqref{lambda pert}. These are valid up to $\beta\mathcal{J}\sim 1/\s$ and so do not agree with with the numerical computations beyond this point.}\label{fig:large_q_integrable}
\end{figure}

Instead of the perturbative solution, we directly solve \eqref{eq: large_q_qtilde_eom} numerically and then compute both the Lanczos coefficients and the Lyapunov exponent. The first step is to compute $g(\tau)$ numerically using a shooting method to solve \eqref{eq: large_q_qtilde_eom}, with thermal boundary conditions $g(0)=g(\beta)=0$. Then, using the equation of motion \eqref{eq: large_q_qtilde_eom} and derivatives of it, we can efficiently compute the moments of $g$, without any additional input other than the numerical value of $g(\tau = \beta/2)$. In this way, we can obtain up to around 16 Lanczos coefficients in the order of seconds and around 35 if one is patient. We fit the slope of the largest 3 even coefficients computed to obtain $\lambda_K$. To compute the Lyapunov exponent we follow the method described in \cite{Jiang:2019pam}, that does not need the explicit form of $g(\tau)$. In Fig.~\ref{fig:large_q_integrable} we plot both $\lambda_K$ and $\lambda_L$ for $\s^2=0.01$ and $\s^2=0.08$, along with the corresponding perturbative Lyapunov exponents \eqref{lambda pert}. We also include in this plot the analytic curve for $\lambda_L = \lambda_K$ in the undeformed model with $\mathfrak{s}^2=0$, see \eqref{large_q_krylov}.

\noindent \textbf{The Lyapunov exponent.} We first note that the numerically computed Lyapunov exponents agree with the perturbative results only up to inverse temperatures $\beta \mathcal{J} \sim 1/\s$, as expected from the perturbative analysis. At larger inverse temperatures, the two results disagree so we cannot draw further conclusions from the perturbative analysis. In particular, we see that the value at which $\lambda_L^{\text{(pert.)}}$ hits zero \cite{Garcia-Garcia:2017bkg}, is not an accurate description of the inverse temperatures in which the actual Lyapunov exponent becomes very small.

Next we note that for fixed $\s^2>0$, the numerically computed Lyapunov exponent initially increases with $\beta\mathcal{J}$ in line with the Lyapunov exponent of $\s^2=0$ before reaching a maximum and then decreasing towards zero. Recall that in section \ref{sec:deformed_finite_q_ch_int}, we observed that at finite $q$ and large $\beta \mathcal{J}$ there is a power-law behaviour at large inverse temperatures that seems $q$ and $s$ dependent. In the regimes analysed here and in \cite{Kim:2020mho}, the power seems to increase with $q$. 
Here we observe a much steeper decay towards zero, that does not seem to fit a power law, maybe in line with the large-$q$ limit taken. It seems that the Lyapunov exponent is reaching zero at a finite temperature in this case. As far as our numerics can assess, the lowest values of $\lambda_L$ we managed to compute are as small as $\sim 10^{-6}$ for both values of $\mathfrak{s}$ analysed. It remains an interesting open question to understand if there is an actual chaotic-to-integrable phase transition at finite temperature in this model.

\

\noindent \textbf{The Krylov exponent.} For fixed $\s^2>0$, the Krylov exponent also initially increases with $\beta\mathcal{J}$ in line with the Krylov exponent of $\s^2=0$, with this behaviour continuing beyond the point that the corresponding Lyapunov exponent reaches its maximum. As in previous cases, we do not observe any non-monotonic behaviour for the Krylov exponent along the flow.

At around the inverse temperatures where the corresponding Lyapunov exponent reaches zero however, the Krylov exponent seems to sharply deviate from the $\s^2=0$ curve. In Fig.~\ref{fig:large_q_integrable} we indicate the first two points that are observed to deviate significantly from the $\s^2=0$ curve by filled circles. Upon examining the Lanczos coefficients from and beyond this point we see significant staggering between even and odd coefficients and it is no longer clear that either the even or the odd coefficients grow linearly. Therefore, these points should \textbf{not} be taken as valid Krylov exponents.

\begin{figure}[H]
        \centering
         \subfigure[$\beta\mathcal{J}=10$]{
                \includegraphics[scale=0.55]{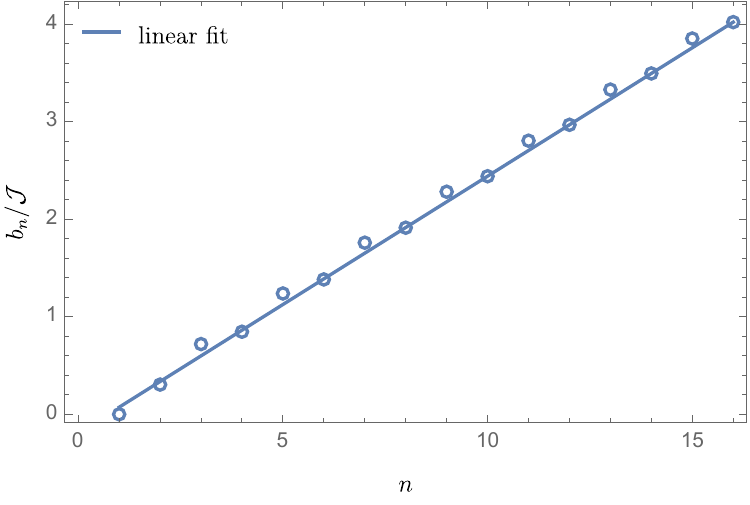}\label{coeffs10}}  \quad\quad
       \subfigure[$\beta\mathcal{J}=100$]{
                \includegraphics[scale=0.55]{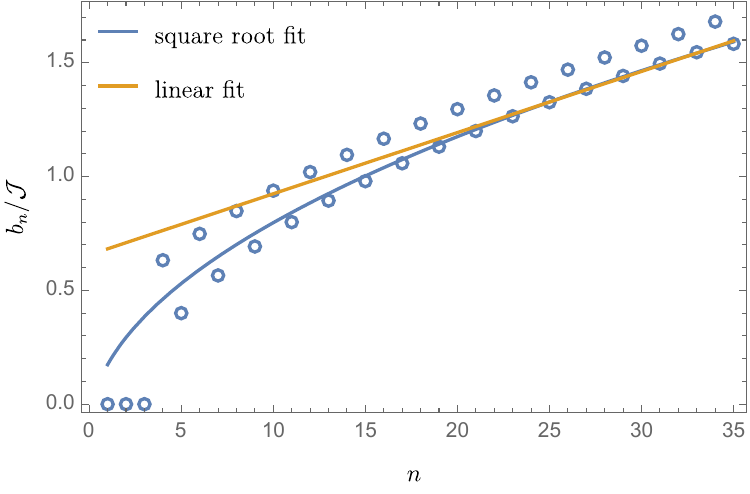} \label{coeffs100}}   
                
                \caption{The Lanczos cofficients for $\s^2=0.08$ at (a) $\beta\mathcal{J}=10$ and (b) $\beta\mathcal{J}=100$. The late coefficients for $\beta\mathcal{J}=10$ have a clear linear fit,  $b_n/\mathcal{J} = 0.26\,n -0.19$. At $\beta\mathcal{J}=100$, the staggering is more pronounced. The late odd coefficients fit with both a square root fit,  $b_n/\mathcal{J} = 0.29 \sqrt{n} -0.11$ and a linear one, $b_n/\mathcal{J} = 0.026\,n+0.65$. Though not shown here, the late even coefficients can also be fit with both linear and square root behaviour.}\label{fig:chaos_integrable_coeffs}
\end{figure}

As evidence, in Fig.~\ref{fig:chaos_integrable_coeffs} we contrast the Lanczos coefficients for $\s^2 = 0.08$ at $\beta\mathcal{J}=10$ and $\beta\mathcal{J}=100$. For the latter case we computed an additional 19 coefficients to ensure we are capturing the asymptotic behaviour. Whilst at $\beta\mathcal{J}=10$, the coefficients clearly grow linearly, at $\beta\mathcal{J}=100$ it is possible to fit both linear and square root growth to the later coefficients (after taking into account the staggering of even and odd coefficients). 

It is suggestive that the Lanczos coefficients stop being conclusively linear at approximately the same inverse temperatures that the Lyapunov exponent becomes close to zero.\footnote{For the finite $q$ to $\tilde q=2$ examples of the previous subsection, the Lyapunov exponent follows instead a power law fall-off. This could be why we do not see the Lanczos coefficients stopping to be linear at finite $q$.} As far as our analysis reaches, we observe monotonic behaviour of the Krylov exponent up to temperatures where we cannot guarantee anymore the existence of a Krylov exponent. The interplay between Krylov methods and chaos-to-integrable phase transitions at finite temperature remains an interesting problem that deserves further investigation. We comment on this in the next section.


\section{Outlook: a conjecture on the growth of the Krylov exponent} \label{sec:outlook}

In this work, we analysed signatures of chaos in a set of SYK-like models in the large $N$ limit, focussing on the Krylov complexity. We started by studying the single SYK model. In the large-$q$ limit, it was already known that the Krylov exponent, $\lambda_K$, is exactly equal to the Lyapunov exponent $\lambda_L$, not only at infinite but also at finite temperature \cite{Parker:2018yvk}. We computed the slope of the Lanczos coefficients to the next order in the $1/q$ expansion, and also numerically at finite-$q$. We found small deviations between the Krylov and Lyapunov exponents in the large $q$ expansion. These deviations were of the order of $\frac{1}{q \, \beta \mathcal{J}}$ for small temperatures. Small differences at finite $q$ were observed numerically, but they fell within our numerical accuracy. The numerical results therefore presented a relatively good match between the Krylov and Lyapunov exponents, both at finite and infinite temperature for finite values of $q$. All the above results reinforced some already existing evidence that the inequality $\lambda_L \leq \lambda_K$ provides a rather tight bound to the Lyapunov exponent which is significantly more restrictive than the chaos bound \eqref{BCHAOS}, namely $\lambda_L \leq 2\pi/\beta$.

This motivated us to study renormalisation group flows in deformed SYK models, for which the Lyapunov exponent was known to have a non-monotonic behaviour, interpolating between two near-maximally chaotic regions or even decreasing to almost zero at very low temperatures when the deformation was integrable. We called these two deformed SYK models, chaos-to-chaos and chaos-to-integrable flows, respectively. With a combination of analytical and numerical techniques we managed to compute the Krylov exponent $\lambda_K$ and the Lyapunov exponent $\lambda_L$ in a variety of these flows both at infinite and finite-$q$.

The results are conclusive for all cases in which the Krylov exponent can be defined. First, the bound $\lambda_L \leq \lambda_K \leq 2\pi/\beta$ is always satisfied. Second, however, the computation of the Krylov exponent fails to capture the non-monotonic behaviour of the Lyapunov exponent. In particular, in cases where the deformation was integrable and $\lambda_L \to 0$ at low temperatures,  the Krylov exponent shows maximal chaos $\lambda_K \to 2\pi/\beta$. We found only one model, the deformed SYK with large $q$ and $\tilde{q}=2$, in which the Lanczos coefficients stop being linear at large $n$. We observe staggering and no definite asymptotic behaviour. It is interesting to note that this behaviour starts happening at inverse temperatures where the Lyapunov exponent becomes very small (or presumably, zero).

It would be desirable to further study this model to assess whether indeed there is a phase transition. If so, one would like to see whether the Krylov complexity stops being exponential and what is its characteristic behaviour at late times. Throughout this work we always assumed, following \cite{Parker:2018yvk}, that Krylov complexity $C_K$ grows exponentially with time and that $\lambda_K$ is given by $2\alpha$, $\alpha$ being the slope of the Lanczos coefficients at large $n$. One should compute the actual Krylov complexity to verify this is actually the case and understand what happens when there is a phase transition. We leave this for future work. 

More generally, one could ask how generic are these non-monotonic behaviours of the Lyapunov exponent. It turns out that for large enough $q$, it is possible to consider a large number of relevant deformations to the single SYK model, namely, by considering Hamiltonians of the form,
\begin{equation}
    H_{\text{def}} = H_q + \sum_i s_i H_{q_i}\,, \qquad s_i \in \mathbb{R} \,,
\end{equation}
where all $q_i$ are different and $q_i < q$ for every $i$. One could imagine parametrically separating the $s_i$ so that the behaviour of the Lyapunov is almost oscillatory as you move to lower temperatures or even engineer a theory that has a particular non-maximal Lyapunov behaviour in the infrared. For the case of the thermal entropy, this has been shown to be possible, at least close enough to the first intermediate near fixed point, both at infinite and finite but large $N$ \cite{unpublished}. It is also possible to relax the Hermiticity condition on the Hamiltonian and allow for complex couplings $s_i$. The generic behaviour of the Krylov exponent in these cases is not known, but see, for instance, related studies of Krylov complexity in open quantum systems \cite{Bhattacharya:2022gbz, Liu:2022god,Bhattacharjee:2022lzy,NSSrivatsa:2023pby}.

Another class of  exactly solvable deformations of SYK Hamiltonians has been studied in \cite{Gross:2019uxi,Gross:2019ach}. These  consist of maps of the original SYK Hamiltonian $H \rightarrow f(H)$. These deformations can be irrelevant, non-local and analytically tractable and can be seen as a quantum mechanical analogue of $T\bar T$ deformations in two-dimensional quantum field theories \cite{Smirnov:2016lqw,Cavaglia:2016oda}. Since they are amenable to analytic treatment, \eg in terms of their two point functions, some results about their Krylov complexity can be found in the literature \cite{He:2022ryk}. It would be interesting to look for a precise connection between these models and the deformed SYK models, if such a connection exists.

Another interesting direction is to check whether the properties of RG flows of SYK models studied in this paper using two different SYK Hamiltonians with a different number of interacting fermions can be connected to the study of non-equilibrium phenomena triggered by a change of the Hamiltonian using those two same Hamiltonians, such as the quench protocol of  \cite{Eberlein:2017wah}.\footnote{Krylov and spread complexity have been studied in time-dependent situations for other systems, see \eg \cite{Nizami:2023dkf,Nizami:2024ltk}.}

\

Going back to the results of this work, it seems tempting to conjecture that (as long as it is well-defined) the Krylov exponent has a monotonic behaviour along thermal RG flow. Namely, in quantum systems that obey unitary evolution at finite temperature,
\begin{eqnarray}
  \beta\partial_\beta \left(\frac{\lambda_{K} \beta}{2\pi}\right) \geq 0 \,. 
\end{eqnarray}
This conjecture is satisfied in all the examples considered in this manuscript where the Krylov exponent could be computed, including the single SYK model. It would be interesting to test this conjecture in other models that exhibit chaos-to-integrable transitions, \eg \cite{Berkooz:2021ehv, Berkooz:2022dfr,Rabinovici:2021qqt,Rabinovici:2022beu,Berkooz:2024evs,Berkooz:2024ofm}. In particular, it would be nice to see whether this conjecture holds at finite $N$, where the Lyapunov exponent has been computed up to $N \sim 60$ using Krylov methods for the single SYK \cite{Kobrin:2020xms} and exact diagonalisation is available up to around $N \sim 34$. Krylov complexity computations for the single SYK model have also been developed at finite $N$ \cite{Jian:2020qpp,Rabinovici:2020ryf}, so it would be desirable to adapt them to include these deformed models. See \cite{Menzler:2024atb} for some progress in this direction. At finite $N$, other quantities that probe late quantum chaos (such as the spectral form factor \cite{Cotler:2016fpe}) could also be used to characterise deformed SYK models, see \cite{Garcia-Garcia:2017bkg} for some progress in this direction. In particular, the relation between the infrared sector that exhibits seemingly non-maximal chaos and holography seems like an interesting question to be explored \cite{Anninos:2018svg,Yoon:2019cql,Jensen:2019cmr}. Conjectures on operator growth have been violated when tested in quantum field theory \cite{Dymarsky:2021bjq,Avdoshkin:2022xuw,Kundu:2023hbk,He:2024xjp}, so studying Krylov complexity in RG flows in QFT might be a natural avenue to test these ideas.

If the conjecture is true and the Krylov exponent is indeed monotonic, then, it would seem that $\lambda_K$ behaves more as an entropy or a $c$-function, rather than as a quantum signature of chaos. In particular, it would mean that $\lambda_K$ is not good at diagnosing chaos in quantum systems where the Lyapunov exponent has non-monotonic behaviour as a function of the energy scale, like, for instance, in systems where the Lyapunov exponent goes to zero at very low energies. See however the discussion in appendix \ref{app:enpassant} which suggests that the slope of early Lanczos coefficients could potentially capture additional properties of such flows. We further provided a concrete example where the Krylov exponent cannot be well-defined after some given inverse temperature, which poses the question of what is the generic behaviour of Krylov complexity (or even what is the simplest signature to assess chaos) in a chaos-to-integrable phase transition in quantum systems.

\

\section*{Acknowledgements}

It is a pleasure to thank Dionysios Anninos, Igal Arav, Micha Berkooz, Mike Blake, Xiangyu Cao, Pawel Caputa, Anatoly Dymarsky, Diego Hofman, Rishabh Jha, Pratik Nandy, Diego Pardo Santos, Lucas S\'a, Adri\'an S\'anchez-Garrido, Edgar Shaghoulian, Joan Sim\'on, Ritam Sinha, David Vegh and Curt von Keyserlingk for interesting discussions. We specially thank Xiangyu Cao and Pratik Nandy for illuminating comments and correspondence on how to compute the Krylov exponent using the pole method. 
The work of SC, SD and OS is supported
by the Israel Science Foundation (grant No. 1417/21), by the German Research Foundation
through a German-Israeli Project Cooperation (DIP) grant “Holography and the Swampland”,
by Carole and Marcus Weinstein through the BGU Presidential Faculty Recruitment Fund
and by the ISF Center of Excellence for theoretical high energy physics. 
The work of DAG is funded by UKRI Stephen Hawking Fellowship ``Quantum Emergence of an Expanding Universe". DAG is further funded by STFC Consolidated grant ST/X000753/1.
SD is supported by an Azrieli fellowship funded by the Azrieli foundation. SC and SD would like to thank the organizers
and participants of the workshop ``Quantum chaos and holography 24" in W\"urzburg for hospitality and for illuminating  discussions. SD gratefully acknowledges the COST Action CA22113 ``THEORY-CHALLENGES'' for supporting a scientific visit to King's College London during which part of this work was carried out.

\appendix


\section{Lyapunov exponent in the deformed SYK}\label{sec: comp Lyapunov}

In this appendix we discuss the numerical method used to find the Lyapunov exponent for the deformed SYK model at finite $q>2$ and large $q$. In both cases we will be concerned with the regularised OTOC defined by
\begin{equation}
\label{regularised OTOC}
    \text{OTOC}\left(t_1,t_2\right) =  \frac{1}{N^2}\sum_{i,j=1}^{N}\text{Tr}\left(\rho^{\frac{1}{4}} \psi_{i}(t_1) \rho^{\frac{1}{4}} \psi_{j}(0) \rho^{\frac{1}{4}} \psi_{i}(t_2) \rho^{\frac{1}{4}} \psi_{j}(0)\right), \quad \rho = \frac{1}{Z(\beta)}e^{-\beta H}~.
\end{equation}
Note that it has be shown that for the SYK, the Lyapunov exponent does not depend on the choice of regularisation \cite{Kobrin:2020xms, Romero-Bermudez:2019vej}. At large $N$ we can write the OTOC as 
\begin{equation}
\label{OTOC large N}
    \text{OTOC}(t_1,t_2) = F_0(t_1,t_2) + \frac{1}{N}F(t_1,t_2)+\cdots~.
\end{equation}
The $1/N$ contribution to the OTOC is described by a set of ladder diagrams that satisfy the equation \cite{Maldacena:2016hyu}
\begin{equation}
\label{Kernel equation}
    F\left(t_1, t_2\right)=\int d t_3 d t_4\; K\left(t_1, t_2, t_3, t_4\right) F\left(t_3, t_4\right)~,
\end{equation}
where the Kernel $K\left(t_1, t_2, t_3, t_4\right)$ is given by
\begin{equation}
\label{kernel}
    K\left(t_1, t_2, t_3, t_4\right)=G^R(t_{13}) G^R(t_{24})\mathcal{J}^2\left(\frac{2^{q-1}}{q}(q-1) G^W(t_{34})^{q-2}+s^2 \frac{2^{\tilde{q}-1}}{\tilde{q}}(\tilde{q}-1) G^W(t_{34})^{\tilde{q}-2}\right)~,
\end{equation}
where $t_{ij} \equiv t_i - t_j$.
Here $G^R(t)$ is the retarded propagator and $G^{W}(t)$ is the Wightman propagator defined between two real time folds separated by half the thermal circle. These are defined by the following relations\footnote{As in the main text, we interpret the expressions $\lim_{\varepsilon\rightarrow 0}G(it\pm\varepsilon)$ as analytically continuing the positive  $\tau>0$ (negative $\tau<0$) branch of the Euclidean correlator (see \eg \cite{Maldacena:2016hyu}), as opposed to naively plugging $it\pm\varepsilon$ in the correlator and performing the limit. An extensive discussion of the different real and imaginary time thermal correlators and the various analytic continuations connecting them can be found in many references, see \eg \cite{Bellac:2011kqa,Negele:1988vy,Sa:2023hmz}.}
\begin{eqnarray}
\begin{cases}
 G^R(t) = \frac{1}{N}\sum_{i}\theta(t)\langle\psi_i(t)\psi_i(0) + \psi_i(0)\psi_i(t)\rangle_{\beta} = \theta(t)\left(G(it + \varepsilon) - G(it-\varepsilon)\right)~, \\
 G^{W}(t) = G(\beta/2 + it)~. \label{def:wightman}
\end{cases}
\end{eqnarray}
\subsection*{Finite \texorpdfstring{$q$}{Lg}}
\begin{figure}[H]
        \centering
         \subfigure[$G^{R}(t)$]{
                \includegraphics[scale=0.5]{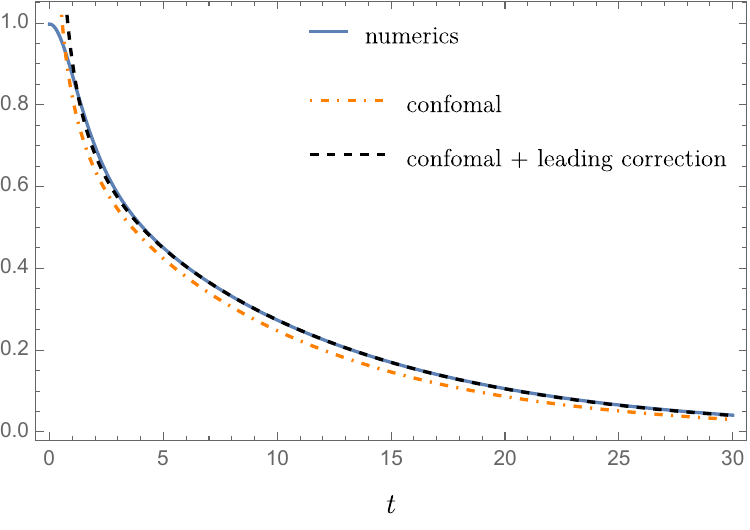}\label{GR_q6}}  \quad\quad
       \subfigure[$G^{W}(t)$]{
                \includegraphics[scale=0.5]{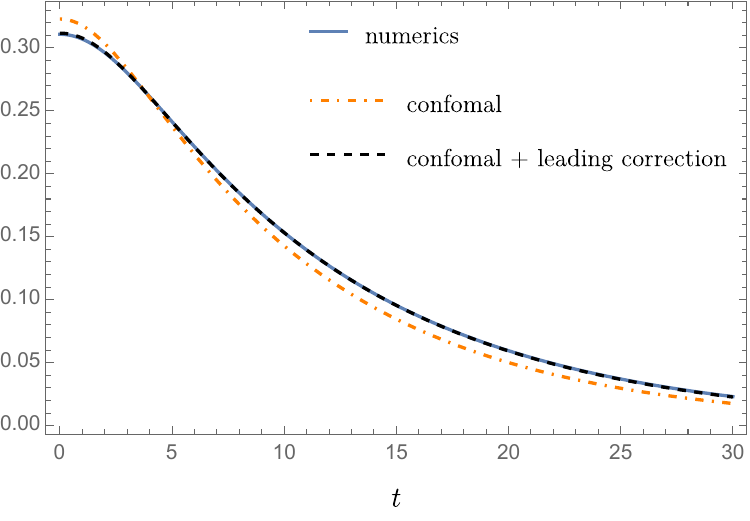} \label{GW_q6}}   
                
                \caption{Numerical computations of $G^{R}(t)$ and $G^{W}(t)$ for $q=6$ at inverse temperature $\beta\mathcal{J}=10$. The numerics (solid lines) are plotted against the conformal solution (dashed-dotted lines) and the conformal solution with leading correction (dashed lines).}\label{fig:GR_GW_q6}
\end{figure}

To find the Lyapunov exponent at finite $q$ we first make the growth ansatz
\begin{equation}
\label{growth ansatz}
  F(t_1,t_2) = e^{\lambda_L(t_1+t_2)/2}f(t_{12})~.
\end{equation}
Writing \eqref{Kernel equation} in frequency space gives
\begin{equation}
    \label{matrix equation}
    f\left(\omega^{\prime}\right) =  \int d\omega \; M(\omega',\omega) f(\omega),
\end{equation}
where
\begin{eqnarray}
\begin{cases}
M(\omega',\omega) &= \frac{\mathcal{J}^2}{2\pi} \left|G^R\left(\omega^{\prime}+i \frac{\lambda_L}{2}\right)\right|^2 \;m(\omega',\omega)~,\\
m(\omega',\omega) &= \int dt\; e^{i \left(\omega^{\prime}-\omega\right) t} \left(\frac{2^{q-1}}{q}(q-1) G^W(t)^{q-2}+s^2\frac{2^{\tilde{q}-1}}{\tilde{q}}(\tilde{q}-1) G^W(t)^{\tilde{q}-2}\right)~.
\end{cases}
\end{eqnarray}
We can find the Lyapunov exponent numerically by discretising $\omega$ and treating \eqref{matrix equation} as a matrix equation. We then search for a value of $\lambda_L$ such that $M(\omega',\omega)$ has an eigenvalue of 1. We do this using binary search to find the value of $\lambda_L$ for which the largest eigenvalue of $M$ crosses 1. To compute the matrix $M(\omega',\omega)$ we first numerically compute the spectral function $\rho(\omega)$ defined by \eqref{eq:spectral_fct_main}.
The numerical computation of $\rho(\omega)$ is given in appendix \ref{app:spectral_fct}. From $\rho(\omega)$ we can find $G^{R}(t)$ and $G^{W}(t)$ by the following relations
\begin{eqnarray}
\begin{cases}
    G^{R}(t) &=  \theta(t)\int d\omega\;\rho(\omega)\cos(\omega t)~, \label{GR}\\ 
    G^{W}(t) &= \int d \omega\; e^{-\omega\left(it+\frac{\beta}{2}\right)} \frac{\rho(\omega)}{1+e^{-\beta \omega}} ~.\label{GW}
\end{cases}
\end{eqnarray}
Note that in order to arrive to this form, one needs to use the fact that $\rho(\omega)$ is even for fermionic systems. 
We can then compute $G^R\left(\omega^{\prime}+i \frac{\lambda_L}{2}\right)$ from  \eqref{GR} by performing a Fourier transform with frequencies shifted by $i\lambda_L/2$. For the single SYK one can check the numerical computation of $ G^{R}(t)$ and $G^{W}(t)$ by comparing to their conformal solutions and leading corrections given by \cite{Maldacena:2016hyu}, 
\begin{eqnarray}
\begin{cases}
    G_{C}^{R}(t) &= 2 \, b \cos (\pi \Delta) \theta(t)\left(\frac{\pi}{\beta\mathcal{J} \sinh \frac{\pi t}{\beta}}\right)^{2 \Delta}\left(1-\frac{\alpha_G(q)}{{\beta\mathcal{J} }} \left(2-\frac{\pi  \tan \left(\pi \Delta\right)+\frac{2 \pi  t}{\beta }}{\tanh \left(\frac{\pi  t}{\beta }\right)}\right)\right)~, \label{GRC}\\ 
    G_{C}^{W}(t) &= b\left(\frac{\pi}{\beta\mathcal{J} \cosh \frac{\pi t}{\beta}}\right)^{2 \Delta} \left( 1 -\frac{\alpha_G(q)}{{\beta \mathcal{J}}} \left(2-\frac{2 \pi  t}{\beta}\tanh \left(\frac{\pi  t}{\beta }\right)\right) \right) ~,\label{GWC}
\end{cases}
\end{eqnarray}
where $\Delta=1/q$, $\alpha_G(q)$ is a $q$ dependent constant that must be fitted numerically and\footnote{Note that to relate this with the conventions of \cite{Maldacena:2016hyu} we used $J^2=\frac{2^{q-1}}{q} \mathcal{J}^2$.}
\begin{equation}\label{constant b}
    b = \frac{1}{2} \left(\frac{(1-2 \Delta ) \tan (\pi  \Delta )}{\pi  \Delta }\right)^{\Delta }~.
\end{equation}
In Fig.~\ref{fig:GR_GW_q6} we show plots with numerical computations of $G^{R}(t)$ and $G^{W}(t)$ against their respective conformal solutions \eqref{GWC} for $q=6$ at inverse temperature $\beta\mathcal{J}=10$. For these plots we use the numerical value of $\alpha_G(6)=0.1737$ given in \cite{Maldacena:2016hyu}.

\subsection*{Large \texorpdfstring{$q$}{Lg}}

Using \eqref{def:wightman}  along with \eqref{large q 1st order} we find that to leading order in the large $q$ limit the kernel \eqref{kernel} is given by
\begin{equation}
K(t_1,t_2,t_3,t_4) = 2\mathcal{J}^2\theta(t_{13})\theta(t_{24})\left(e^{g(\beta/2+it)}+ s^2 e^{\frac{g(\beta/2+it)}{\mathfrak{n}}}\right)~.
\end{equation}
Substituting this into \eqref{Kernel equation} with the growth ansatz \eqref{growth ansatz} and applying $\partial_{t_1}\partial_{t_2}$ to both sides of the equation gives 
\begin{equation}
\label{lyapunov diff eqn1}
\left(\frac{\lambda_L}{4}-\partial_t^{2}\right)f(t) = 2\mathcal{J}^2\left(e^{g(\beta/2+it)}+ s^2 e^{\frac{g(\beta/2+it)}{\mathfrak{n}}}\right)f(t)~.
\end{equation}
We can then find $\lambda_L$ by looking for a normalisable solution to this equation. This can be done numerically by solving the equation with the Runge-Kutta method starting at $t=0$ and going up to some large value of $t$. From \eqref{regularised OTOC} and \eqref{OTOC large N} we see that $F(t_1,t_2)=F(t_2,t_1)$, which in turn means that $f(t_{12})$ is an even function. This gives us the initial condition $f'(0) = 0$. The initial value of $f(0)$ can be any positive real number for the procedure since the value of the Lyapunov exponent does not depend on the overall normalisation of $f$. We then do a binary search for $\lambda_L$ looking for the solution to cross zero at the end of the interval. To compute the plots in Fig.~\ref{fig:lyapunov} we used initial conditions $f(0)=0.01, f'(0)=0$ and applied the Runge-Kutta method on the interval $t \in [0, 3\beta]$.


\section{Spectral function in the deformed SYK}\label{app:spectral_fct}

In this appendix we describe the numerical method used to compute the spectral function of $G(t)$, labelled by $\rho(\omega)$ and defined by \eqref{eq:spectral_fct_main}. The method we use follows that described in \cite{Sa:2021tdr,Sa:2023hmz} for an  SYK Lindbladian model, though it is also applicable for unitary deformed SYK models. The procedure involves numerically solving a system of equations involving $\rho(\omega)$ and the analogously defined spectral function of $\Sigma(t)$, which we label by $\sigma(\omega)$. Specifically, after passing to real time and Fourier transforming (see supplementary materials of \cite{Sa:2021tdr} for a detailed derivation), the Schwinger-Dyson equations \eqref{SD2} can be written as\footnote{Our $\rho(\omega)$ and $\sigma(\omega)$ are $\rho^-(\omega)$ and $\sigma^-(\omega)$, respectively, in the conventions of \cite{Sa:2021tdr}.}
\begin{eqnarray}
\begin{cases}
    \rho(\omega)&=\frac{\sigma(\omega)}{(\omega +\pi \sigma^H(\omega))^2+(\pi \sigma(\omega))^2}~,\\
    \sigma(\omega)&=\mathcal{J}^2\left(\frac{2\cosh\left(\beta\omega/2\right)}{q}\tilde{\rho}^{*(q-1)}(\omega)+s^2 \frac{2\cosh\left(\beta\omega/2\right)}{\tilde{q}}\tilde{\rho}^{*(\tilde{q}-1)}(\omega)\right)~,\label{fourier SD}
\end{cases}
\end{eqnarray}
where $\sigma^H$ is the Hilbert transform, 
\begin{equation}
\label{sigmaH}
	 \sigma^H(\omega) = -\frac{1}{\pi}\mathcal P\int \mathrm d \nu \frac{\sigma(\nu)}{\omega-\nu}~,
\end{equation}
with $\mathcal{P}$ denoting the Cauchy principal value. The spectral function $\tilde{\rho}(\omega)$ is given by $\tilde{\rho}(\omega)\equiv\rho(\omega)/\cosh(\beta\omega/2)$, while $\tilde{\rho}^{*(n)}(\omega)$ denotes the $n$-fold convolution defined by
\begin{equation}
\label{convolution}
	\tilde{\rho}^{*(n)}(\omega) = \int \left[\prod_{i=1}^{n-1}\mathrm d\mu_i\right] \,\tilde{\rho}(\omega-\sum_{i=1}^{n-1} \mu_i)\prod_{i=1}^{n-1}\tilde{\rho}(\mu_i)~.
\end{equation}
The equations \eqref{fourier SD} are then solved numerically on a discretised frequency grid. The size and spacing of the grid needed depends on how quickly the spectral function $\rho(\omega)$ decays, which in turn depends the values of $q$ and the inverse temperature $\beta\mathcal{J}$. For our purposes we typically used a grid spanning from $\omega=-30$ to $\omega=30$ in steps of $\Delta\omega=0.01$. 

The numerical procedure begins with an ansatz for $\rho(\omega)$ which we take to be $\rho(\omega)\sim 1/(\omega^2+0.1)$. At each step in the iteration $\sigma(\omega)$ is computed from $\rho(\omega)$ using the second equation of \eqref{fourier SD} and then $\sigma^{H}(\omega)$ is computed from \eqref{sigmaH}. Finally, $\rho(\omega)$ is updated by a proportion of the error in the first equation of \eqref{fourier SD},
\begin{equation}
\rho_{j+1}(\omega) = \rho_{j}(\omega) + \eta \left(\frac{\sigma_{j}(\omega)}{(\omega +\pi \sigma_{j}^H(\omega))^2+(\pi \sigma_{j}(\omega))^2}-\rho_{j}(\omega)\right)~,
\end{equation}
where the weight $\eta$ is initially  set to 0.5. After each iteration we keep track of the difference
\begin{equation}
\rho_{\text{diff}}=\sum_{\omega}|\rho_{j+1}(\omega)-\rho_{i}(\omega)|~.
\end{equation}
If the difference increases we drop the value of the weighting parameter $\eta$ by half. We terminate the procedure when $\rho_{\text{diff}}<10^{-8}$. The $n$-fold convolution \eqref{convolution} can be carried out extremely efficiently in \verb|Mathematica| using the built-in  \verb|ListConvolve| function. Due to the appearance of diverging factors of $\cosh(\beta\omega/2)$, a further trick is needed to avoid large numbers causing numerical instabilities when $\beta\mathcal{J}\gtrsim 10$. The trick involves expanding out the factors of $\cosh(\beta\omega/2)$ multiplying the convolutions in the second equation of \eqref{fourier SD} in such a way that the resulting equation consists of convolutions only involving $\rho(\omega)$ and $\rho(\omega)\tanh(\beta\omega/2)$. For example, if $q=4$ we would write the factor of $\cosh(\beta\omega/2)$ as
\begin{equation}
\label{trigExpand}
\cosh(\beta\omega/2) = \cosh\left(\beta\left(\omega-\mu_1-\mu_2\right)/2 + \beta\mu_1/2 +\beta\mu_2/2\right)~,
\end{equation}
and then expand the three terms in the argument of the RHS using the addition formula. This results in four terms involving products of hyperbolic cosines and hyperbolic sines, that simplify with the hyperbolic cosines in the denominator of the convolution $\tilde{\rho}^{*(3)}(\omega)$. This trick can be trivially extended to any value of $q$. The procedure converges in the order of seconds on a standard laptop.

\begin{figure}[H]
        \centering
         \subfigure[$\rho(\omega)$]{
                \includegraphics[scale=0.5]{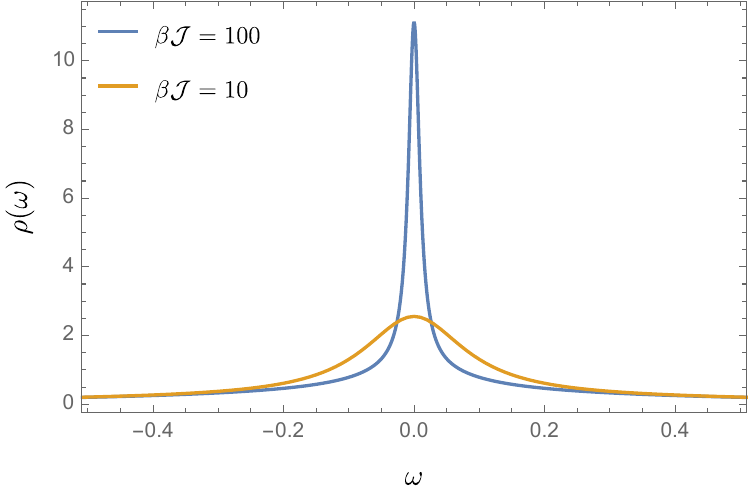}\label{rho}}  \quad\quad
       \subfigure[$\sigma(\omega)$]{
                \includegraphics[scale=0.5]{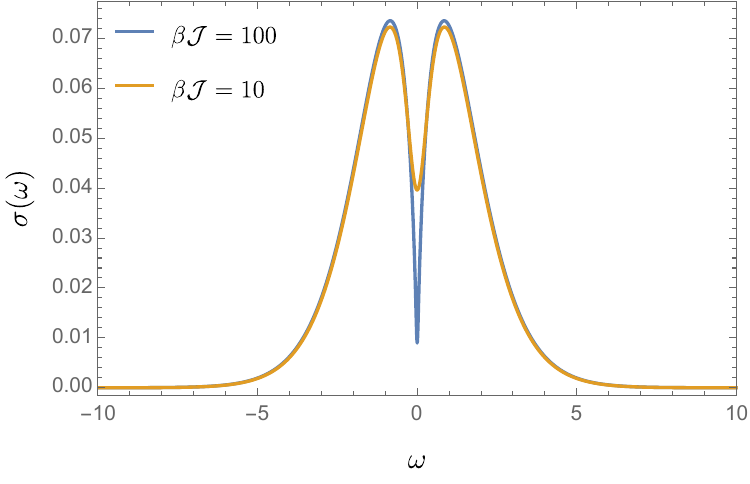} \label{sigma}}   
                
                \caption{Spectral functions computed for the single SYK model with $q=6$ at inverse temperatures $\beta\mathcal{J}=10,100$. For large inverse temperatures the spectral functions are more sharply peaked around the origin.}\label{fig:rho_sigma}
\end{figure}

In Fig.~\ref{fig:rho_sigma} we plot numerical computations of $\rho(\omega)$ and $\sigma(\omega)$ for the single SYK model with $q=6$ at inverse temperatures of $\beta\mathcal{J}=10,100$. We note that as $\beta\mathcal{J}$ increases the spectral functions become more sharply peaked around the origin, making it more difficult to compute them accurately. Using $\rho(\omega)$ computed with this method we were able to compute the Lyapunov and Krylov exponents for inverse temperatures of up to $\beta\mathcal{J}\sim 125$.

\section{Krylov exponent from moments in the deformed SYK}\label{app:bump}

In section \ref{sec:Chaos-to-chaos flows} we computed the Krylov exponent at large-$q$ for the deformed SYK with $q=2\tilde{q}$. This was done both by computing Lanczos coefficients as described in section \ref{sec: lanczos from moments}, and from the pole of the autocorrelation function as described in section \ref{sec: slope from pole}. In this case, with the moments method we are limited by numerical errors from the algorithm \eqref{coeffs from moments} to computing a maximum of $17$ Lanczos coefficients and $\lambda_{K}$ is computed from the slope of the last $3$. 

In the results shown in Fig.~\ref{fig:lyapunov} we observe that for the first method there is a small deviation from $\lambda_{K}(\beta/2\pi)=1$ during the transition between the two regions of near-maximal chaos which is not seen from the pole of the autocorrelation function. This happens, for instance, around $\beta\mathcal{J}\sim10^{7.5}$ for $s=0.001$. 

\begin{figure}[H]
        \centering
         \subfigure[]{
                \includegraphics[scale=0.5]{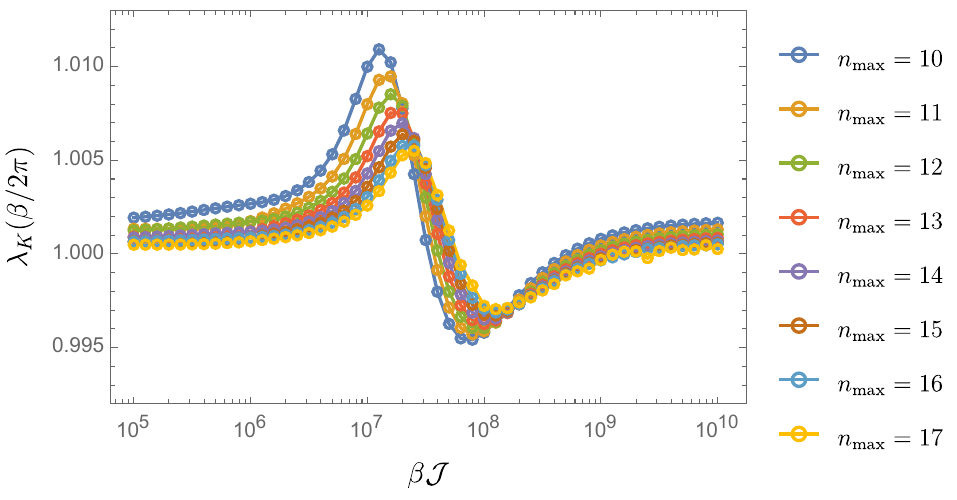}\label{fig:bump}}  \quad\quad
       \subfigure[]{
                \includegraphics[scale=0.5]{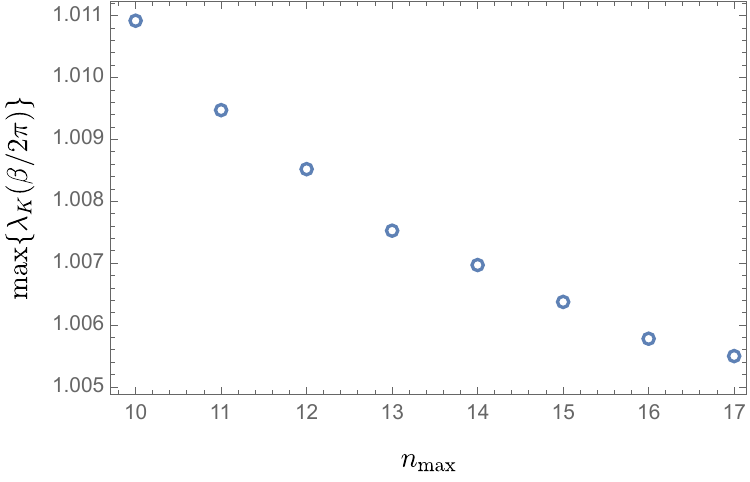} \label{bump_height}}   
                \caption{(a) The Krylov exponent, computed from the slope of the Lanczos coefficients, in the large $q$ deformed model with $q/\tilde{q}=2$ and $s=0.001$. Each curve corresponds to a different number of coefficients computed, $n_{\text{max}}$, before taking the slope of the last $3$. (b) The maximum value of $\lambda_{K}(\beta/2\pi)$ in plot (a) for each value of $n_{\text{max}}$.}\label{fig:bump_height}
\end{figure}

We believe the deviation observed to be an artefact of not computing enough Lanczos coefficients before taking the slope. To demonstrate this in Fig.~\ref{fig:bump} we plot the Krylov exponent for the flow with $s=0.001$ and different values of the number of Lanczos coefficients computed, $n_{\text{max}}$. In each case $\lambda_K$ is computed from the last $3$ coefficients. We observe that as we increase $n_{\text{max}}$ the deviation decreases. This is also shown in Fig.~\ref{bump_height} which plots the maximum value of $\lambda_{K}(\beta/2\pi)$. We expect this trend to continue and that if enough coefficients were computed one would no longer observe a significant deviation from $\lambda_{K}(\beta/2\pi)=1$ in this region, in agreement with the result from the pole of the autocorrelation function.

\section{The slope of the first Lanczos coefficients in the deformed SYK} \label{app:enpassant}

\begin{figure}[H]
        \centering
         \subfigure[$s=1$]{
                \includegraphics[scale=0.45]{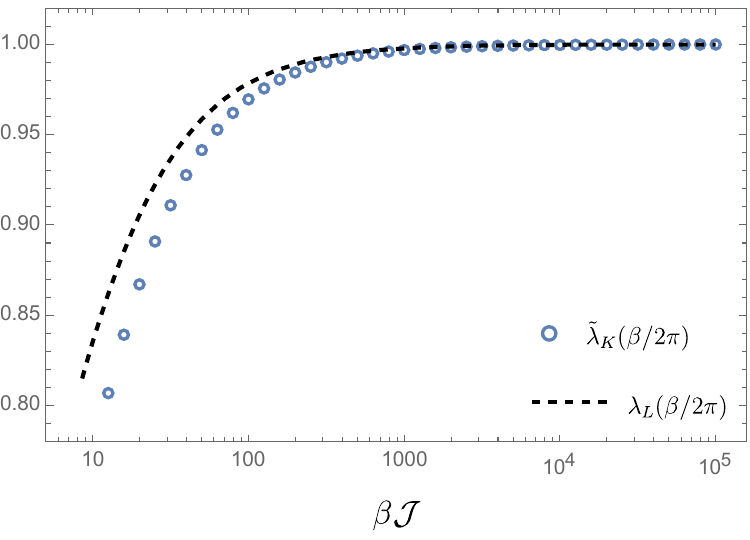}\label{early_s1}}  \quad\quad
       \subfigure[$s=0.1$]{
                \includegraphics[scale=0.45]{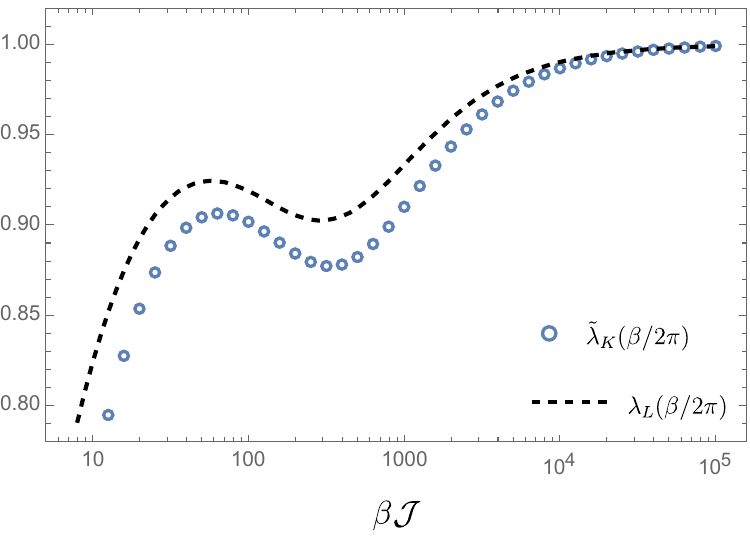} \label{early_s0p1}}   
       \subfigure[$s=0.01$]{
                \includegraphics[scale=0.45]{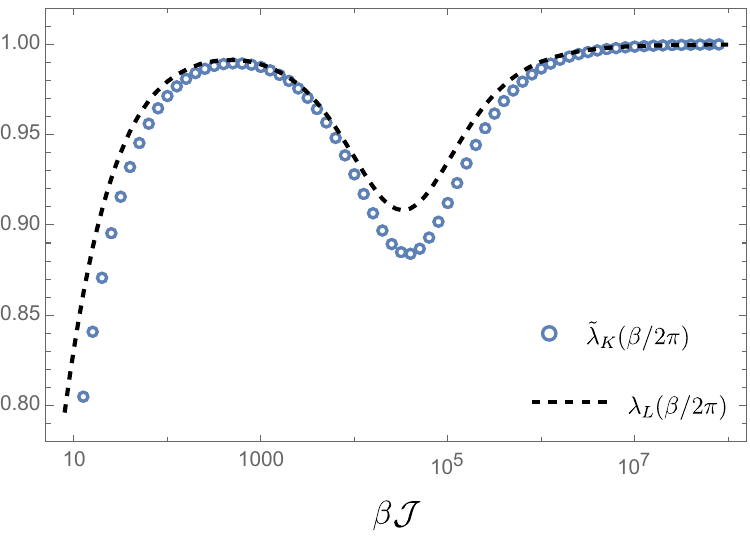}\label{early_s0p01}}  \quad\quad
       \subfigure[$s=0.001$]{
                \includegraphics[scale=0.45]{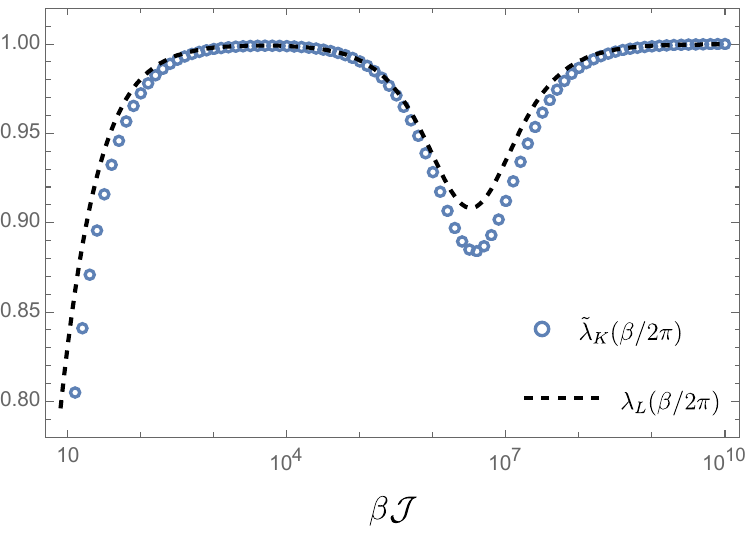} \label{early_s0p001}}                           
                \caption{$\tilde{\lambda}_{K}$ and Lyapunov exponent as a function of $\beta\mathcal{J}$ for the deformed model \eqref{deformed Hamiltonian} in the large-$q$ limit with $q=2\tilde{q}$. The blue dots are computed from the first two Lanczos coefficients which are found by the moments method described in section \ref{sec: lanczos from moments}. The black dashed lines show the Lyapunov exponent, which is computed numerically. Plots are shown for different values of $s$.}\label{fig:early-lyapunov}
\end{figure}

In section \ref{sec:Chaos-to-chaos flows} it is shown that for the deformed SYK model at large-$q$ with $q=2\tilde{q}$, the Krylov exponent does not detect the non-monotonic behaviour and sub-maximal chaos that the Lyapunov exponent shows between the two regimes of maximal chaos, see Fig.~\ref{fig:lyapunov}. If however, instead of looking at the asymptotic growth of the Lanczos coefficients at large $n$, we focus on the first few coefficients, we do observe a non-monotonic behaviour in this region. Surprisingly, up to an overall shift, the slope of just the first two Lanczos coefficients can be used to define a quantity that closely resembles the behaviour of the Lyapunov exponent. This is shown in Fig.~\ref{fig:early-lyapunov}, where we compute
\begin{equation}
\tilde{\lambda}_{K} = b_2-b_1 - \left(\max_{\beta\mathcal{J}}\{b_2-b_1\}-\frac{2\pi}{\beta}\right)~,
\end{equation}
and compare to the Lyapunov exponent for the same set of deformed SYK models as shown in Fig.~\ref{fig:lyapunov}. Though the match is not exact, it would be interesting to understand why this crude measure seems to be able to probe the region of sub-maximal chaos while $\lambda_{K}$ is not.

\bibliographystyle{JHEP}
\bibliography{bibliography}

\end{document}